\documentclass[11pt]{article}
\pdfoutput=1
\usepackage{amsmath,amssymb}
\usepackage{graphicx,color}

\numberwithin{equation}{section}

\def\({\left(}
\def\){\right)}

\newcommand{\be}{\begin{equation}}
\newcommand{\ba}{\begin{eqnarray}}
\newcommand{\ea}{\end{eqnarray}}
\newcommand{\ee}{\end{equation}}

\begin{document}

\begin{titlepage}
\thispagestyle{empty}


\vspace{.4cm}
\begin{center}
\noindent{\Large \textbf{A perturbative study on the analytic continuation for generalized gravitational entropy}}\\
\vspace{2cm}

Andrea Prudenziati

\vspace{1cm}
  {\it
Institute of Physics, University of $S\tilde{a}o$ Paulo \\
05314-970 $S\tilde{a}o$ Paulo, Brazil\\
\vspace{0.2cm}
\vskip 1ex
{\tt prude@if.usp.br }
 }

\vskip 2em
\end{center}

\vspace{.5cm}
\begin{abstract}
We study the analytic continuation used by Lewkowycz and Maldacena to prove the Ryu-Takayanagi formula for entanglement entropy, which is the holographic dual of the trace of the $\beta$-power of the time evolution operator when $\beta\in \mathbb{R}$. This will be done perturbatively by using a weakly time dependent Hamiltonian, corresponding to a small shift of the dual static background. Depending on the periodicity we impose on the gravitational solution, we consider two different possibilities and compare the associated entropies with the results obtained through a minimal area computation. To our surprise we discover that, at first order, both choices correctly reproduce the associated entanglement entropy. Furthermore we find unexpected divergent contributions that we have to discard in order to fit the minimal area computation, and an additional requirement that needs to be imposed on the $\beta$ dependence on the metric.
\end{abstract}

\end{titlepage}

\newpage


\newpage

\tableofcontents

\section{Introduction}

Entanglement Entropy has been an intensively studied subject in the last few years, and indeed a primary role has been played by the well known formula for its holographic computation. Given a fixed time codimension one submanifold $A$, with its complement $B$, and a density matrix $\sigma$, the entanglement entropy is defined to be the Von Neumann entropy for the reduced density matrix $\sigma_A=Tr_{H_B}[\sigma]$, where $H_A$ ( resp. $H_B$ ) is the Hilbert space living on $A$ ( resp. $B$ ): $S_A=-Tr_{H_A}[\sigma_A \log{\sigma_A}]$. Then, if we have a dual static Euclidean background, \cite{Ryu:2006bv} proposed the formula
\begin{equation}\label{for}
S_A=\frac{\mathcal{A}_{min}}{4 G_N} ,
\end{equation}
where $G_N$ is the Newton constant in the holographic $d+1$ dimensional space $\mathcal{M}$, and $\mathcal{A}_{min}$ is the minimal surface area inside $\mathcal{M}$, extending towards $\partial \mathcal{M}$ where the boundary $\partial\mathcal{A}_{min}$ is located, and such that $\partial\mathcal{A}_{min}=\partial A$.

Not only this remarkable formula provides a relatively simple computational tool for $S_A$, so far only possible at weak coupling using the replica trick for path integration\footnote{see also a recent field theory perturbative computation \cite{Rosenhaus:2014woa}}, it also allowed many new developments in studying the properties of entanglement entropy. We may mention for example an alternative proof of the c-theorem and possible higher dimensional generalizations \cite{Casini:2012ei}; numerous achievements in reconstructing bulk and gravity properties from field theory data, for instance \cite{Faulkner:2013ica} \cite{Lashkari:2013koa} \cite{Spillane:2013mca}  and \cite{Swingle:2014uza}, in particular enlightening the connection between gravity equations of motion and a sort of thermodynamic first law for entanglement entropy \cite{Bhattacharya:2012mi} ( extended to a zeroth and second law in \cite{Allahbakhshi:2013rda} ). Further the Ryu-Takayanagi formula may be used to construct an incredibly simple proof of strong subadditivity \cite{Headrick:2007km}, although its counterpart for time dependent backgrounds looks more tricky and deeply interconnected with the null energy condition \cite{Wall:2012uf}, and for instance \cite{Allais:2011ys} \cite{Callan:2012ip} for the case of Vaidya space.

Even if many consistency checks have been provided, a proof for this simple formula was so far lacking and has been searched for a few years; a first proposal by Fursaev \cite{Fursaev:2006ih} was later discovered by \cite{Headrick:2010zt} to use gravitational solutions with conical singularities. Later a proof was given at $d=2$ $AdS_3/CFT_2$, by \cite{Faulkner:2013yia} and \cite{Hartman:2013mia}, and for generic dimension but $A$ restricted to a sphere by \cite{Casini:2011kv}. Only recently the complete generalization was provided by \cite{Lewkowycz:2013nqa}, through the introduction of the concept of generalized gravitational entropy ( see \cite{Fursaev:2014tpa} for some comments ).

The proposed proof develops from the holographic correspondence between $Tr\left[\rho^{\beta}\right]$, with $\rho$ the non normalized time evolution operator $\rho=\mathcal{T}e^{-\int_0^{2\pi}H(t)}$, $H(t)=H(t+2\pi)$, and the exponential of a certain gravitational action $S^{grav}_{Eucl}(\beta)$ 
\begin{equation}\label{ass}
Tr\left[\rho^{\beta}\right]=e^{-S^{grav}_{Eucl}(\beta)} ,
\end{equation}
with the following properties: first the new, regular background $\mathcal{M}_{\beta}$, on which the action is evaluated, contains a circle noncontractible on $\partial\mathcal{M}_{\beta}$, along which a Euclidean time coordinate $t\in[0,2\beta\pi)$ runs. Second, when $\beta\in\mathbb{Z}$, there are two periodicity conditions to be imposed on the fields $\phi$ entering $S^{grav}_{Eucl}(\beta)$: $\phi(t) = \phi(t+2\pi)$ and $\phi(t) = \phi(t+2\beta\pi)$  \footnote{periodicity in $t \sim t+2\pi$ needs a priory to be defined only on $\partial\mathcal{M}_{\beta}$. We will however extend it to the full bulk.}. Third, if $\beta$ is analytically continued from integers to reals, so that the two periodic conditions are no longer compatible, only $t \sim t+2\pi$ is preserved and the time integral for the action is redefined as $\int_0^{2\beta\pi}dt\rightarrow \beta\int_0^{2\pi}dt$. Then we can compute the entropy for $\tilde{\rho}\equiv\rho/Tr[\rho]$ as $S=-Tr[\tilde{\rho}\log{\tilde{\rho}}]=\partial_{\beta}S^{grav}_{Eucl}(\beta)|_{\beta=1}-S^{grav}_{Eucl}(1)$. 

Now let us restrict to the boundary $\partial\mathcal{M}_{\beta}$, and consider a codimension one submanifold $A$ at $t_E=const$ inside $\partial\mathcal{M}_{\beta}$, with $t$ a different coordinate chosen to wind around ( not along ) $\partial{A}$. Locally $t$ is the angle obtained by transforming $t_E$ and $x_1$ to polar coordinates $t,r$, where $t_E,x_1$ span the space perpendicular to $\partial{A}$. As we wind around multiple times we are naturally implementing the replica trick used for the field theory computation of $S_A$, that is instead of path integrating over a $\beta$-fold cover of the boundary space with cuts along $A$, we compactify the time direction and go around $\beta$-times, with $A$ as a Cauchy surface on $\partial\mathcal{M}_{\beta}$. When we look for its holographic counterpart $\mathcal{M}_{\beta}$, we are led to the construction described in the above paragraph. 

\begin{figure}[h]
\centering
\vspace{-0pt}
\includegraphics[width=0.9\textwidth]{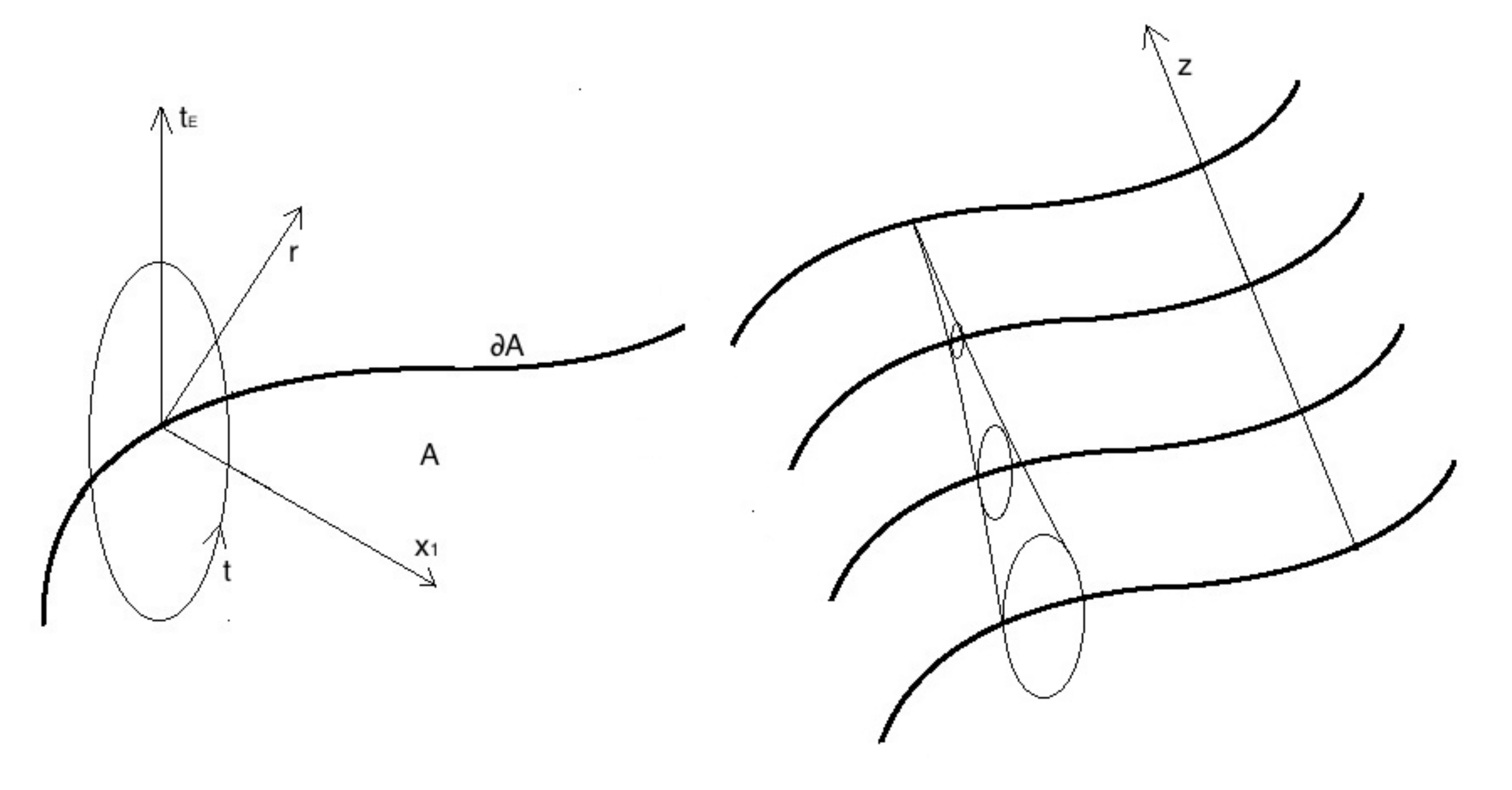}
\vspace{-0pt}
\caption{On the left we have a representation of the coordinates $t,r$ and $t_E,x_1$ on $\partial\mathcal{M}_{\beta}$. On the right we show the radial additional direction of $\mathcal{M}_{\beta}$ and the eventual shrinking of the circle parameterized by $t$, somewhere in the interior; because of its definition, contributes to $S$ come only from there.}
\label{figura0}
\end{figure}

 Thus the Ryu-Takayanagi formula can be translated to a purely gravitational computation:
\begin{equation}\label{ass2}
\partial_{\beta}S^{grav}_{Eucl}(\beta)|_{\beta=1}-S^{grav}_{Eucl}(1)=\frac{\mathcal{A}_{min}}{4 G_N}.
\end{equation}

One might wonder were does the time dependence on the left hand side of the above (\ref{ass2}) come from, if on the right hand side we are considering a static background? In polar coordinates $t,r$, the part of the regular metric at the boundary of $\partial\mathcal{M}_{\beta}$ spanning the plane perpendicular to $\partial A$, may be chosen to  locally look like
\[
ds_{\perp}=dr^2+\frac{r^2}{\beta^2}dt^2 .
\]
To avoid the proper length of the $t$-circle to shrink at zero at $r=0$, we need to perform a bulk coordinate transformation such to induce a Weyl rescaling on $\partial\mathcal{M}_{\beta}$ corresponding to a local multiplication by $\sim r^2$. If the shape of $\partial A$ is generic, this can be done only at the price of introducing a time dependence in the rescaled metric. 

The intent of the present paper is to study in detail and explicitly check the assumption, made in \cite{Lewkowycz:2013nqa}, regarding the analytic continuation of $\beta$ for the gravitational solution $\mathcal{M}_{\beta}$, used to compute $S^{grav}_{Eucl}(\beta)$. Although reasonable, it is in fact not the unique choice that could have been made, and we therefore think it is compelling to better understand it. One can of course follow a more pragmatic point of view and accept it by the a posteriori justification that it works, but this would then spoil the status of proof of the full construction.

We will work perturbatively in the time dependent deformation of the static bulk metric \footnote{another work that has studied a quite related problem from a perturbative perspective, but restricted on the boundary, is \cite{Lewkowycz:2014jia}}. A second goal is then to construct a formalism for comparing minimal area to generalized gravitational entropies at first order, with particular care on the cutoff procedure ( that has so far been quite neglected in the literature ) and on the possible periodicity conditions for the gravitational solution. These techniques could then be applied to more general theories where the equivalent of a minimal area formula is not known, for example in order to study the dynamics of the entanglement entropy in these cases; we will discuss it further in the conclusions.

Consider a weakly time dependent Hamiltonian
\begin{equation}\label{hami}
H(t)=H_0+\delta h(t)  \;\;\;\;\; \delta<<1 .
\end{equation}
As we have discussed below equation \ref{ass2}, this would correspond to $A$  being a submanifold only marginally different from either half of the AdS boundary or a sphere, cases for which the rescaled metric remains $t-$independent. The problem is that this change of shape for $\partial A$ makes difficult to compute minimal surfaces, even at first order in the deformation. Another possibility for introducing weak time dependence is to consider the  perturbed AdS metric, with perturbations satisfying the linearized equations of motion ( and being in general time dependent ), and $A$ remaining untouched. Our $\mathcal{M}_{\beta}$ will then be AdS + first order perturbations, the nontrivial $\beta$ dependence contained inside these last ones and fixed by the periodicity conditions along the time direction that we will choose. We want to compare the generalized gravitational entropy on this background computed using different choices for the analytic continuation in $\beta$, with the results from minimal area computation on the same background at $\beta=1$. Note that, even if for time dependent Lorentzian manifolds the recipe is to look for extremal surfaces instead of minimal, \cite{Hubeny:2007xt}, as we are at first order the shape of the extremal surface remains the same as in the unperturbed AdS, and it then coincides with the minimal one, while its area will obviously be modified. 

The strategy will be as follows. First we will consider the analytic continuation in $\beta$ of $Tr\left[\rho^{\beta}\right]$ from a purely field theoretical point of view, expanding the Hamiltonian as in (\ref{hami}) in order to show that, in principle, different possibilities arise and how to select the correct one. This will be done in section \ref{one}. In section \ref{two} we introduce the bulk gravitational version of the same problem, representing the Euclidean time evolution operator holographically as in (\ref{ass}), and show what kind of additional complications arise in trying to select the correct choice when $\beta$ becomes real. To warm up and set the technique we will start with simple unperturbed AdS in section \ref{sec1} ( that is $H=H_0$ ), matching the minimal area with the generalized gravitational entropy result as in (\ref{ass2}). This will be done by selecting once and for all the simple case where $A$ is just half of the boundary; in particular we will see the nontrivial role played by the UV cutoffs that has not been covered in detail by \cite{Lewkowycz:2013nqa}. Understood the static case, for which the analytic continuation in $\beta$ is a trivial problem,  we will start to tackle the time dependent case by perturbing the AdS metric  or, equivalently, by switching on $ h(t) $ in the Hamiltonian. In section \ref{aaa} we set some common results and, finally, in sections \ref{bbb} and \ref{ccc} we consider two different possibilities for the analytic continuation of the gravitational solution. The first choice will be selected by requiring the gravity solution to preserve periodicity after a time shift of $2\pi\beta$, when $\beta$ becomes real, and thus giving up periodicity for time shifts of $2\pi$. This is different by what have been used in \cite{Lewkowycz:2013nqa}; the Lewkowycz-Maldacena choice instead ( preserving $2\pi$-periodicity and giving up the $2\pi \beta$-one) will be considered in the last section. Both results will be compared to what we obtain using the Ryu-Takayanagi formula, which is done at $\beta=1$ and is thus independent by whatever choice we made for the analytic continuation. Finally we present our conclusions. 

\section{Analytic continuation from a field theory perspective}\label{one}

For the boundary field theory how to define the analytic continuation of $\beta$ in $Tr\left[\rho^{\beta}\right]$ is in principle clear: first we define the modular Hamiltonian as minus the logarithm of $\rho$:
\begin{equation}\label{dens}
\rho=\mathcal{T}e^{-\int_0^{2\pi}H(t)}\equiv e^{-\mathcal{H}} .
\end{equation}
Then, simply
\begin{equation}\label{corret}
\rho^{\beta}=e^{-\beta\mathcal{H}}  \;\;\;\;\; \beta\in \mathbb{R} .
\end{equation}
Thus ( with $\tilde{\rho}$ the normalized $\rho$ ) the entropy computed from the replica trick
\[
S=-\partial_{\beta}\log{Tr\left[\rho^{\beta}\right]}|_{\beta=1}+\log{Tr\left[\rho\right]}=-\partial_{\beta}\log{Tr \left[e^{-\beta\mathcal{H}}\right]}|_{\beta=1}+\log{Tr\left[ e^{-\mathcal{H}}\right]}=
\]
\begin{equation}\label{master}
=\frac{Tr \left[\mathcal{H} e^{-\mathcal{H}}\right]}{Tr\left[ e^{-\mathcal{H}}\right]}+\log{Tr\left[ e^{-\mathcal{H}}\right]}
\end{equation}
correctly coincides with
\[
S=-Tr\left[\tilde{\rho}\log{\tilde{\rho}}\right]=-\frac{Tr\left[\rho\log{\rho}\right]}{Tr [\rho]}+\log{Tr [\rho]}=\frac{Tr \left[\mathcal{H} e^{-\mathcal{H}}\right]}{Tr\left[ e^{-\mathcal{H}}\right]}+\log{Tr\left[ e^{-\mathcal{H}}\right]} .
\]
The problem is that, in general, the exact form for the modular Hamiltonian is not known. What is achievable is an order by order expansion formula called Magnus expansion:
\begin{equation}\label{magnus}
\mathcal{H}=\int_0^{2\pi}dt_1\;H(t_1)-\frac{1}{2}\int_0^{2\pi}dt_1\int_0^{t_1}dt_2 \;[H(t_1),H(t_2)] 
\end{equation}
\[
+\frac{1}{6}\int_0^{2\pi}dt_1\int_0^{t_1}dt_2\int_0^{t_2}dt_3 \;\left([H(t_1),[H(t_2),H(t_3)]]+[[H(t_1),H(t_2)],H(t_3)]\right)+\dots .
\]
We consider the case 
\[
H(t)=H_0+\delta h(t)  \;\;\;\;\; \delta<<1
\]
and $H_0$ time independent. The goal is to compute $Tr \left[\mathcal{H} e^{-\mathcal{H}}\right]$ at first order in $\delta$. The computation is done in Appendix (\ref{C}) and it gives 
\begin{equation}\label{ham}
Tr \left[\mathcal{H} e^{-\mathcal{H}}\right]_{order \; O(\delta)}=Tr \left[\mathcal{T}\left(\int_{0}^{2\pi}dt\;H(t) e^{-\int_{0}^{2\pi}d\tilde{t}\;H(\tilde{t})}\right)\right]_{order \;O(\delta)} .
\end{equation}
The above formula agrees with the entanglement entropy variation computed in \cite{Rosenhaus:2014woa} for a generic normalized density matrix $\rho=\rho_0+\delta\rho$, and reproducing the first law of \cite{Bhattacharya:2012mi} \cite{Blanco:2013joa}, when $\rho$ is taken to be (\ref{dens}) and 
\[
\delta\rho=-\int_0^{2\pi}h(t)\;e^{-2\pi H_0}
\]
with $Tr{\rho}=Tr{\rho_0}=Tr{\delta\rho}=0$ \footnote{I thank Aitor Lewkowycz for this comment}.

Now suppose we didn't know about the modular Hamiltonian and of (\ref{corret}), and we were looking for the analytic continuation of $Tr \rho^{\beta}$ from the very definition of $\rho$. One possible guess would be to use the equality valid for $\beta\in \mathbb{Z}$ and periodic Hamiltonian $H(t+2\pi)=H(t)$, 
\begin{equation}\label{unob}
Tr\left[\left(\mathcal{T}e^{-\int_0^{2\pi}H(t)}\right)^{\beta}\right]=Tr\left[\mathcal{T}e^{-\int_0^{2\beta\pi}H(t)}\right]   \;\;\;\;\; \beta\in \mathbb{Z},\;\;\;\;\; H(t+2\pi)=H(t),
\end{equation}
and extend it to $\beta\in \mathbb{R}$. This however fails for real $\beta$ as
\[
-\partial_{\beta} Tr\left[\mathcal{T}e^{-\int_0^{2\beta\pi}H(t)}\right]_{\beta=1}= Tr\left[ 2\pi \mathcal{T}\left( H(0) e^{-\int_0^{2\pi}H(t)}\right)\right] \neq Tr \left[\mathcal{H} e^{-\mathcal{H}}\right] ,
\]
where the last inequality is already evident at first order in $\delta$. 

A second guess might be
\begin{equation}\label{dueb}
Tr\left[\mathcal{T}e^{-\beta\int_0^{2\pi}H(t)}\right]  \;\;\;\;\;  H(t+2\pi)=H(t) ,
\end{equation}
that correctly reproduces (\ref{ham}) at first order in $\delta$,
\[
-\partial_{\beta} Tr\left[\mathcal{T}e^{-\beta\int_0^{2\pi}H(t)}\right]_{\beta=1}= Tr\left[ \mathcal{T}\left( \int_{0}^{2\pi}dt\; H(t) e^{-\int_0^{2\pi}d\tilde{t}\;H(\tilde{t})}\right)\right] ,
\]
but fails to equate $Tr \rho^{\beta}$ for integer $\beta$ 
\[
Tr\left[\left(\mathcal{T}e^{-\int_0^{2\pi}H(t)}\right)^{\beta}\right]\neq Tr\left[\mathcal{T}e^{-\beta\int_0^{2\pi}H(t)}\right]   \;\;\;\;\; \beta\in \mathbb{Z},\;\;\;\;\; H(t+2\pi)=H(t).
\]
So far we have considered an Hamiltonian that, even in its analytically continued version,  remained $2\pi$ periodic and $\beta$ independent, but other possible choices could involve different periodicities, as for example with $2\beta\pi$ period, and/or nontrivial $\beta$ dependence of the Hamiltonian. The moral then is that the correct way of choosing the analytic continuation is far from trivial and simple guesses may easily turn out to be wrong. Let us now look at what happens when we move to the bulk.

\section{Analytic continuation from a gravitational perspective}\label{two}

We would like to consider now the holographic counterpart of the previous section. 

Given the classical correspondence
\begin{equation}\label{bes}
Tr\left[\left(\mathcal{T}e^{-\int_0^{2\pi}H(t)}\right)^{\beta}\right]=e^{-S_{Eucl}^{grav}(\beta)} , \;\;\;\;\; \beta\in \mathbb{Z}
\end{equation}
the question is how do we analytically continue the gravitational side in order to obtain the dual of $Tr\left[\rho^{\beta}\right]$ for $\beta\in \mathbb{R}$? This time we do not have an analogue of the boundary modular Hamiltonian that we can use; on the other side, as it is a classical problem, all the troubles given by the time ordering disappear.

We will start by considering what periodicity we can impose on the analytically continued action; when $\beta\in \mathbb{R}$ the double periodicity with periods $2\pi$ ( for representing the same operator $\rho$ $\beta$-times ) and $2\beta\pi$ ( for closing the trace ) is no longer possible, except for constant functions, so we have to choose between preserving one or the other. When we keep $L_{Eucl}^{grav}(\beta,t+2\pi)=L_{Eucl}^{grav}(\beta,t)$ it is natural to define the analytic continuation by time integrating between zero and $2\pi$ and then multiplying by $\beta$:
\begin{equation}\label{ddue}
Tr\left[\left(\mathcal{T}e^{-\int_0^{2\pi}H(t)}\right)^{\beta}\right]\leftrightarrow e^{-\beta\int_{0}^{2\pi}dt\;L_{Eucl}^{grav}(\beta,t)}, \;\;\;\;\; \beta\in \mathbb{R} .
\end{equation}
If instead we require $L_{Eucl}^{grav}(\beta,t+2\beta\pi)=L_{Eucl}^{grav}(\beta,t)$ we are lead to the second proposal:
\begin{equation}\label{unno}
Tr\left[\left(\mathcal{T}e^{-\int_0^{2\pi}H(t)}\right)^{\beta}\right]\leftrightarrow e^{-\int_{0}^{2 \beta\pi}dt\;L_{Eucl}^{grav}(\beta,t)}, \;\;\;\;\; \beta\in \mathbb{R} .
\end{equation}
Note that when $\beta$ is restricted to be an integer (\ref{ddue}) automatically becomes periodic also in $t\sim t+2\beta\pi$. Instead the opposite does not happen for  (\ref{unno}), so we need to require as part of the definition, that $L_{Eucl}^{grav}(\beta,t)$ in this case obeys the double periodicity whenever $\beta$ is restricted to be an integer. This poses obvious constraints on the action and makes $\int_{0}^{2 \beta\pi}dt\;L_{Eucl}^{grav}(\beta,t)$ different from the simple rescaling of the integration extrema of $\int_{0}^{2\pi}dt\;L_{Eucl}^{grav}(\beta,t)$. 

We note that (\ref{ddue}) corresponds in fact to the prescription given in \cite{Lewkowycz:2013nqa}. We would then expect it to give the correct result and any other choice to fail. Despite this, and because the final goal is to learn more about the analytic continuation, we will still pursue the computation for both proposals above. Further let us point out that any relationship between analytic continuations in the bulk, as the above (\ref{unno}), (\ref{ddue}), and their possible boundary counterparts is in general complicated. Thus correctly guessing what is the dual of  (\ref{corret}), or (\ref{unob}) and (\ref{dueb}) is not a simple task.

One would love to now repeat the analysis done in the previous section, deriving with respect to $\beta$ and trying to relate the result to the boundary stress-energy tensor. The computation however, although apparently simple, is far from trivial; for (\ref{unno}) we have
\begin{equation}\label{p1}
S=\partial_{\beta}\int_{0}^{2 \beta\pi}dt\;L_{Eucl}^{grav}(\beta,t)|_{\beta=1}-\int_{0}^{2 \pi}dt\;L_{Eucl}^{grav}(1,t)=
\end{equation}
\[
=\int_{0}^{2 \pi}dt\;\partial_{\beta}L_{Eucl}^{grav}(\beta,t)|_{\beta=1}+2\pi L_{Eucl}^{grav}(1,2\pi)-\int_{0}^{2\pi}dt\;L_{Eucl}^{grav}(1,t) ,
\]
while for (\ref{ddue})
\begin{equation}\label{p2}
S=\partial_{\beta}\left(\beta\int_{0}^{2\pi}dt\;L_{Eucl}^{grav}(\beta,t)\right)_{\beta=1}-\int_{0}^{2 \pi}dt\;L_{Eucl}^{grav}(1,t)=\int_{0}^{2 \pi}dt\;\partial_{\beta}L_{Eucl}^{grav}(\beta,t)|_{\beta=1} .
\end{equation}
If the dependence by $\beta$ was only in the metric we could rewrite 
\[
\partial_{\beta}L_{Eucl}^{grav}(\beta,t)=\partial_{\beta}(\gamma_{\mu\nu}(\beta))\frac{\partial}{\gamma_{\mu\nu}(\beta)} L_{Eucl}^{grav}(\beta,t)\sim T_{00} ,
\]
where $\gamma_{\mu\nu}$ is the boundary induced metric ( on the cutoff surfaces we will soon introduce ), and $T_{00}$ the boundary induced stress energy momentum tensor. This because $\frac{\partial}{g_{\mu\nu}(\beta)} L_{Eucl}^{grav}(\beta,t)$ vanishes because of the bulk equations of motion, and the only $\beta$ dependent metric component is $\gamma_{00}$, as we will see. What goes wrong however, apart from the additional terms in (\ref{p1}), is the fact that the very cutoff surface we will use to define our gravitational action, are in general $\beta$ dependent, creating in this way further contributions. Thus we need a more precise computation to check the validity or not of either (\ref{ddue}) and (\ref{unno}) at order $O(\delta)$, in order to correctly reproduce the entanglement entropy through the replica trick.

\section{Comparing Ryu-Takayanagi formula to the generalizad gravitational entropy in unperturbed AdS}\label{sec1}

Let us start from the time independent case, where we consider $\mathcal{M}=AdS_5$ with the submanifold $A$, from now on, being half of the AdS boundary. First we will compute the minimal surface in this case, setting the benchmark against which to check the results obtained from the generalized gravitational entropy computation. Being time independent the various proposals for "$AdS_5(\beta)$" are equivalent; nonetheless we will here set up the computation technique and introduce the cutoff surfaces we will use throwout the paper. 

\subsection{Minimal area}

The Poincare coordinates in the Euclidean signature cover the full AdS space, unlikely in the Lorentzian case where they cover only half. Thus we can use them even for computing a minimal surface area that cuts in two the full AdS,
\begin{equation}\label{g}
ds^2=\frac{R^2}{z^2}\left( dz^2+dt_E^2+dx_1^2+dx_2^2+dx_3^2\right) .
\end{equation}
The area of the minimal surface at $t_E=0$ whose boundary at $z=0$ is $x_1=0$, is simply given by the integration of $\sqrt{\det(\partial_a x^{\mu}\partial_b x^{\nu}g_{\mu\nu})}$, where $a,b$ are $x_2,x_3$ and $z$:
\begin{equation}
S_A=\frac{\mathcal{A}_{min}}{4 G_N}=\frac{1}{4 G_N}\int_{-L_2/2}^{L_2/2} dx_2\int_{-L_3/2}^{L_3/2} dx_3 \int_{\epsilon_z}^{\infty}dz \;\frac{R^3}{z^3}=\frac{L_2 L_3 R^3}{8 G_N \epsilon_z^2} ,
\end{equation}
where $\epsilon_z$ is the UV cutoff and $L_2,L_3$ compactify the $x_2,x_3$ direction.

\subsection{Gravitational action}

We start from the Poincare metric in polar coordinates with period $2\pi\beta$:
\[
 t_E=r\sin(t/\beta) \; \;\;\;\;\; x_1=r\cos(t/\beta) ,
\]
\begin{equation}\label{f}
ds^2=\frac{R^2}{z^2}\left( dz^2+dr^2+\frac{r^2}{\beta^2}dt^2+dx_2^2+dx_3^2\right) .
\end{equation}
This metric is regular everywhere and obviously periodic in $t\sim t+2\pi$ as well. The first problem is to find coordinates such that the proper size of the euclidean time circle doesn't vanish ( as $r\rightarrow 0$ ) at the boundary of AdS, but eventually only in the interior. 
\begin{figure}[h]
\centering
\vspace{-0pt}
\includegraphics[width=0.5\textwidth]{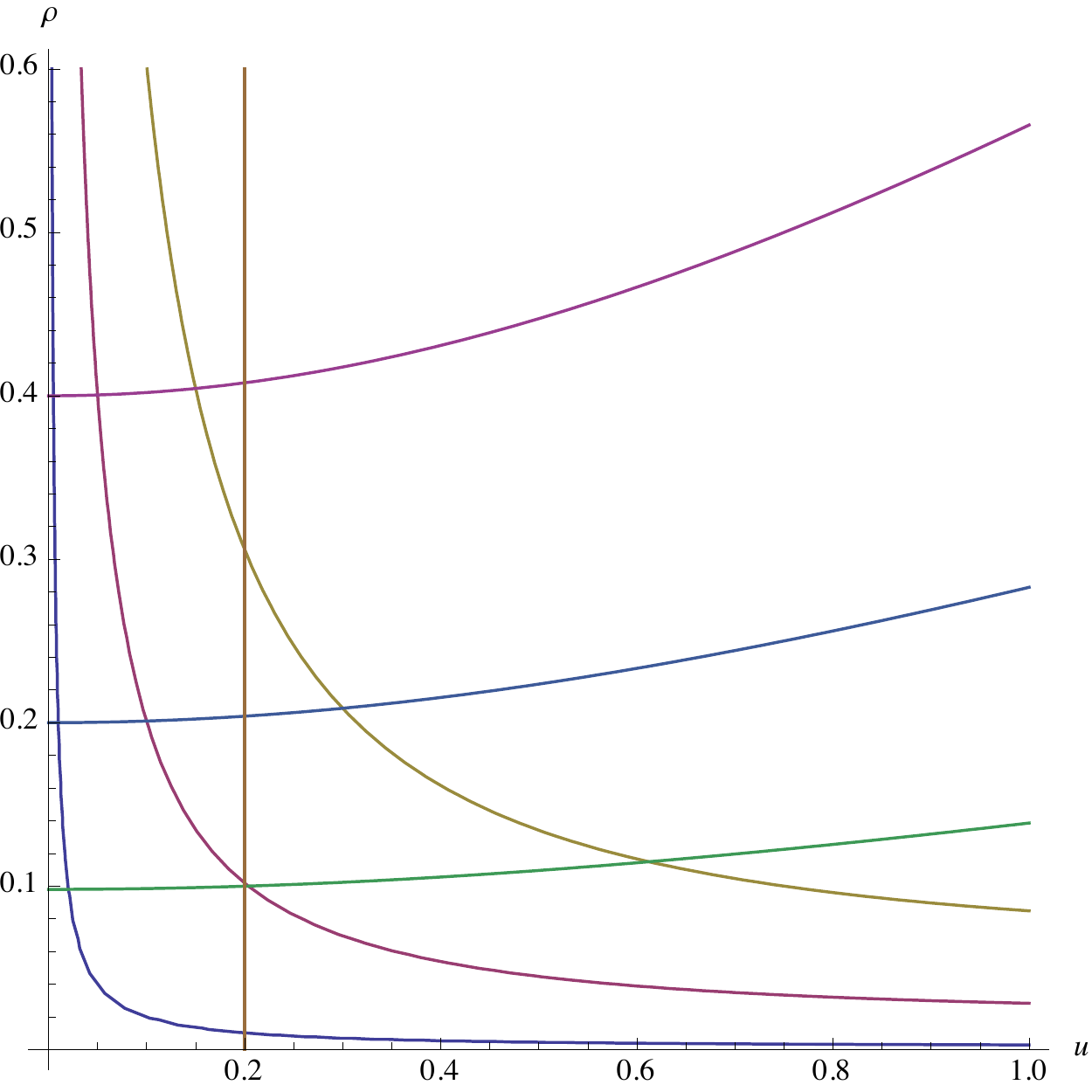}
\vspace{-0pt}
\caption{the hyperbolic lines are $z=$ const, the slightly uprising ones are $r=$ const and the vertical line is the cutoff in $u$. The values displayed are totally random and have no meaning at all.}
\label{figura1}
\end{figure}
Because of this we change the bulk coordinates such that the Weyl rescaled metric at the boundary of AdS has a nonvanishing time circle. One possibility is the following transformation
\begin{equation}
z(\rho,u)=\frac{\rho u}{\beta\sqrt{1+\frac{u^2}{\beta^2}}}  \;\;\;\;\; r(\rho,u)=\frac{\rho }{\sqrt{1+\frac{u^2}{\beta^2}}} ,
\end{equation}
with inverse
\begin{equation}\label{inveq}
u(r,z)=\frac{z \beta}{r}  \;\;\;\;\; \rho(r,z)=\sqrt{r^2+z^2} .
\end{equation}
They produce
\begin{equation}\label{coordan}
ds^2=\frac{R^2}{u^2}\left(\frac{\beta^2}{\beta^2 + u^2} du^2 +dt^2 +\frac{\beta^2+u^2}{\rho^2 }(d\rho ^2 + dx_2^2 + dx_3^2)\right) .
\end{equation}
The gravitational action is
\begin{equation}\label{action}
-S_{Eucl}^{grav} =  \frac{1}{16 \pi G_N}\int_{AdS} \sqrt{g}({\mathcal{R}}+\frac{12}{R^2}) +\frac{1}{8 \pi G_N}\int_{\partial AdS} \sqrt{\gamma}\;\Theta ,
\end{equation}
with $\Theta=\gamma^{\mu\nu}D_{\mu}n_{\nu}$, $n$ the vector normal to $\partial AdS$ pointing outside, that is away from the bulk, and with unit norm, and $\gamma_{\mu\nu}$ the induced metric $\gamma_{\mu\nu}=g_{\mu\nu}-n_{\mu}n_{\nu}$ on the boundary. Note that $\sqrt{\gamma}$ is the determinant on the coordinates of the boundary, while the contraction for $\Theta$ is other all.

For $\partial AdS$ we mean the cutoff surface approaching the boundary. We will also temporarily need to introduce an additional cutoff on $r$, $r=\epsilon_r$; it will be removed at the end of the computation. 
\begin{figure}[h]
\centering
\vspace{-0pt}
\includegraphics[width=0.5\textwidth]{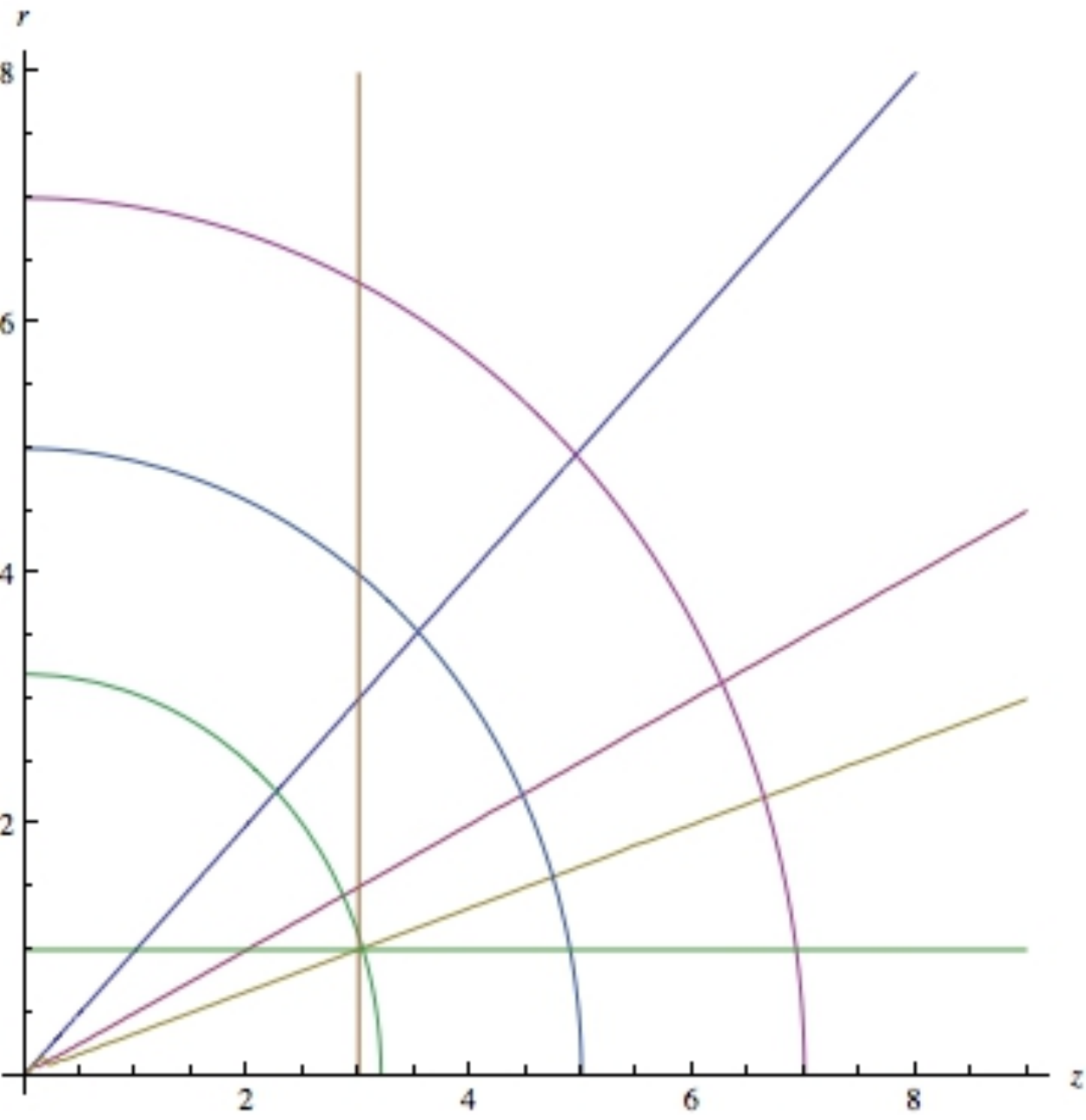}
\vspace{-0pt}
\caption{The straight lines at different angles are $u=$ const, the almost circular ones are $\rho =$ const, while the vertical and horizontal ones are respectively the cutoffs in $z$ and $r$.}
\label{figura2}
\end{figure}

The point of using different coordinates lies entirely in having a different cutoff surface that approaches the AdS boundary, such that the metric there has a nonvanishing circle proper size for the coordinate $t$. This new cutoff is achieved fixing $u=\epsilon_u$, with $\epsilon_u$ some constant. On the other hand the computation in the original Poincare coordinates had a cutoff in $z=\epsilon_z$. The relationship between these two cutoffs is given by the first equation in (\ref{inveq}), and it depends on $r$. As we want $\epsilon_u$ to be constant we need to fix $r$ to some value, effectively changing the cutoff procedure in the computation of the generalized gravitational entropy. Because the minimal area surface will be computed in $r,z$ coordinates, we can still use the cutoff $z=\epsilon_z$ for that computation provided that the two cutoffs agree at the value of $r$ at which the minimal surface is located, that is $r=0$ regularized as $r=\epsilon_r$: 
\begin{equation}\label{cutoff1}
\epsilon_u =\frac{\epsilon_z \beta}{\epsilon_r} .
\end{equation}
The cutoff on $r$ induces an equivalent cutoff on $\rho$, which depends on the coordinate $u$ as well:
\begin{equation}\label{cutoff2}
\epsilon_{\rho}=\epsilon_r\sqrt{1+\frac{u^2}{\beta^2}}.
\end{equation}

We can now compute the action (\ref{action}) using the normal, outward pointing, unit vector to the boundary ( at $u=\frac{\epsilon_z \beta}{\epsilon_r}$ ), which is
\[
n=-\frac{u}{\beta R}\sqrt{\beta^2+u^2}\;\partial_u |_{u=\frac{\epsilon_z \beta}{\epsilon_r}} .
\]
The result is
\begin{equation}\label{actioncom}
-S_{Eucl}^{grav} =  2\pi \beta L_2 L_3 \left(-\int_{\epsilon_u}^{\infty}du \int_{\epsilon_{\rho}}^{\infty}d\rho \frac{\beta R^3 (\beta^2 + u^2)}{2\pi  G_N  \rho^3 u^5} + \int_{\epsilon_{\rho}}^{\infty}d\rho  \frac{R^3 (\beta^2 + u^2)(4 \beta^2 + u^2)}{
 8\pi   G_N\beta  \rho^3 u^4 } |_{u=\epsilon_u}\right)=
\end{equation}
\[
=\frac{(3 \epsilon_r^2 + \epsilon_z^2) L_2 L_3 R^3}{8 G_N \epsilon_z^4 } .
\]
From this expression it is immediate to compute the result in the limit $\epsilon_r \rightarrow 0$:
\begin{equation}
S= -S_{Eucl}^{grav}+\beta\partial_{\beta}S_{Eucl}^{grav} = \frac{ L_2 L_3 R^3}{8 G_N \epsilon_z^2 } ,
\end{equation}
that coincides with the result from the minimal surface\footnote{as we introduced a temporary cutoff $\epsilon_r$ one might wonder if there is a need for a theta term in the action also for this temporary boundary. If introduced however, it is easy to show that the correction to the above formula vanishes in the $\epsilon_r\rightarrow 0$ limit}.

\section{Common results for first order perturbations of AdS}\label{aaa}

Having found agreement between the entanglement entropy computed from the Ryu-Takayanagi formula and the corresponding generalized gravitational entropy when restricted to the static AdS case, we now move to study its variation when the AdS metric has been perturbed. As we have already discussed  we will first select some analytic continuation, then perturb the metric accordingly, solve its equations of motion and use the result to compute the generalized gravitational entropy. Finally we will compare with what has been obtained using  the minimal surface procedure. But before starting we need some general results, independent of the choice for the analytic continuation. 

Naively one would think that, given a generic metric perturbation $h_{\mu\nu}$, $\Delta S_{Eucl}^{grav}=S_{Eucl}^{grav}(h_{\mu\nu}\neq 0)-S_{Eucl}^{grav}(h_{\mu\nu}=0)$ would be zero at first order as the background metric satisfies the equations of motion. In fact the boundary term of (\ref{action}) is there exactly to ensure this when a manifold with boundary is considered. However the caveat is that we work with a cutoff defining a surface on which, as long as the cutoff is not removed, the value of  $h_{\mu\nu}$ is in general not vanishing ( this will be clear when we will solve the equations of motion for $h_{\mu\nu}$ ). And the condition for the $\Theta$ term in (\ref{action}) to kill the boundary term is exactly that the perturbation of the metric has to vanish on the boundary. There are in total four boundary surfaces, two defined by the cutoffs (\ref{cutoff1}) and (\ref{cutoff2}), and two by the corresponding infrared versions $\L_u$ and $L_{\rho}$, needed to regularize the action when in presence of metric perturbations:
\begin{equation}\label{cutoff3}
\L_u =\frac{\L_z \beta}{\epsilon_r} ,
\end{equation}
\begin{equation}\label{cutoff4}
\L_{\rho}=\L_r\sqrt{1+\frac{u^2}{\beta^2}} .
\end{equation}
It may be argued that $\epsilon_{\rho},\;L_{\rho}$ and $\epsilon_u,\;L_u$ are not on the same level, the first two to eventually be removed at the end of the computation, the seconds to be kept to regularize a truly divergent quantity. Sending $\epsilon_r\rightarrow 0$ for example ( or $L_r\rightarrow\infty$ ), removes the corresponding $\epsilon_{\rho}$ ( $L_{\rho}$ ), while it keeps $\epsilon_u,\;L_u$\footnote{ that are both pushed to infinity, restricting the integration to the region in AdS where the proper length of the time circle goes to zero, see figure (\ref{figura0})}. We will always include a $\Theta$ term for the boundaries in $\epsilon_u,\;L_u$, and consider both the case in which a $\Theta$ term has been included for those in $\epsilon_{\rho}$ and $L_{\rho}$, and the one in which it has not. The final result will be independent by the choice ( for simplicity we temporarily indicate both kind of boundaries as $\partial AdS$, even though this is not technically correct ).

From \cite{Wald:1984rg} we have that the variation of the bulk part of the action is
\[
\Delta \left(\frac{1}{16 \pi G_N}\int_{AdS} \sqrt{g}({\mathcal{R}}+\frac{12}{R^2}) \right)= \frac{1}{16 \pi G_N}\int_{\partial AdS} \sqrt{\gamma}\;v_{\mu}n^{\mu}+ eq.\;of\;motion ,
\]
with $n^{\mu}$ the normal vector to the boundary already encountered and ( calling $g_{\mu\nu}$ the AdS unperturbed metric ) 
\[
v_{\mu}n^{\mu}=n^{\mu}g^{\nu\rho}\left(\bigtriangledown_{\rho}h_{\mu\nu}-\bigtriangledown_{\mu}h_{\nu\rho}\right)=n^{\mu}\gamma^{\nu\rho}\left(\bigtriangledown_{\rho}h_{\mu\nu}-\bigtriangledown_{\mu}h_{\nu\rho}\right)
\]
( where the last equality comes from symmetric-antisymmetric tensor contraction ). Moreover
\[
\Delta\left( \frac{1}{8 \pi G_N}\int_{\partial AdS} \sqrt{\gamma}\;\Theta\right)= \frac{1}{8 \pi G_N}\int_{\partial AdS} \left(\Delta\left(\sqrt{\gamma}\right)\;\Theta+\sqrt{\gamma}\; \Delta\left(\Theta\right)\right) , 
\]
with
\[
\sqrt{\gamma}\;\Delta\left(\Theta\right)=\frac{1}{2} n^{\mu}\gamma^{\nu\rho}\bigtriangledown_{\mu}h_{\nu\rho} .
\]
Thus it remains
\begin{equation}\label{m}
-\Delta S_{Eucl}^{grav} =\frac{1}{16 \pi G_N}\int_{\partial AdS}\left(\sqrt{\gamma}n^{\mu}\gamma^{\nu\rho}\bigtriangledown_{\rho}h_{\mu\nu}+2 \Theta\Delta\left(\sqrt{\gamma}\right)\right) .
\end{equation}
Note that both terms would vanish if $h_{\mu\nu}$ was zero on $\partial AdS$.

Without the $\Theta$ term instead we would have
\begin{equation}\label{n}
-\Delta S_{Eucl}^{grav} =\frac{1}{16 \pi G_N}\int_{\partial AdS}\sqrt{\gamma}\;n^{\mu}g^{\nu\rho}\left(\bigtriangledown_{\rho}h_{\mu\nu}-\bigtriangledown_{\mu}h_{\nu\rho}\right) .
\end{equation}
The final goal is to compute the perturbation to the Entanglement Entropy from the formula $\Delta S=\left( -\Delta S_{Eucl}^{grav}+\beta\partial_{\beta}\Delta S_{Eucl}^{grav} \right)_{\beta=1}$. According to the two choices (\ref{unno}) and (\ref{ddue}) we have either (\ref{p1}) or (\ref{p2}) that we conveniently rewrite
\begin{equation}\label{pp1}
\Delta S=-\int_{0}^{2\pi}dt\;\Delta L_{Eucl}^{grav}(1,t)+2\pi \Delta L_{Eucl}^{grav}(1,2\pi)+\int_0^{2\pi} dt \;\partial_{\beta}\left( \Delta  L_{Eucl}^{grav}(\beta,t)\right)_{\beta=1} ,
\end{equation}
\begin{equation}\label{pp2}
\Delta S=\int_0^{2\pi} dt \;\partial_{\beta}\left( \Delta  L_{Eucl}^{grav}(\beta,t)\right)_{\beta=1} .
\end{equation}

 As we already discussed, the dependence on $\beta$ of $\Delta L_{Eucl}^{grav}$ comes both from the metric and the extremes of integration, (\ref{cutoff1}), (\ref{cutoff2}), (\ref{cutoff3}), and (\ref{cutoff4}) . Choosing a boundary orientation we can write
\[
\int_0^{2\pi} dt \;\Delta  L_{Eucl}^{grav}(\beta,t)=
\]
\begin{equation}\label{q}
=\int_0^{2\pi} dt \;\Big( \int_{\epsilon_{\rho}(\beta,\epsilon_u(\beta))}^{L_{\rho}(\beta,\epsilon_u(\beta))}d\rho\;s_{\rho}(\beta,\epsilon_u(\beta),\rho,t)-\int_{\epsilon_{\rho}(\beta,L_u(\beta))}^{L_{\rho}(\beta,L_u(\beta))}d\rho\;s_{\rho}(\beta,L_u(\beta),\rho,t)
\end{equation}
\[
-\int_{\epsilon_u(\beta)}^{L_u(\beta)}du\;s_{u}(\beta,u,\epsilon_{\rho}(\beta,u),t)+\int_{\epsilon_u(\beta)}^{L_u(\beta)}du\;s_{u}(\beta,u,L_{\rho}(\beta,u),t)\Big) ,
\]
where $\epsilon_{\rho}(\beta,\epsilon_u(\beta))$ and $L_{\rho}(\beta,\epsilon_u(\beta))$ refer to the cutoffs (\ref{cutoff2}) and (\ref{cutoff4}) evaluated in $u=\epsilon_u(\beta)$, and analogously for $\epsilon_{\rho}(\beta,L_u(\beta))$ and $L_{\rho}(\beta,L_u(\beta))$. We notice however that the direct $\beta$ dependence of $\epsilon_{\rho}$ and $L_{\rho}$ in fact cancels with the induced one by fixing $u=\epsilon_u(\beta),L_u(\beta)$, as can be checked straightforwardly. Thus for the two integrals in the first line of (\ref{q}), the overall $\beta$ derivative of (\ref{pp1}) and (\ref{pp2}) passes directly on the integrand. Finally $s_{\rho}$ is a short notation for the integrand at fixed time of (\ref{m}) and $s_{u}$ for either (\ref{m}) or (\ref{n}). Relative signs appear in (\ref{q}) as we choose to maintain the same orientation for the perpendicular $n$ vector to both IR and UV boundaries, as explained in Appendix \ref{A}. 

Then we need to solve the equations for $h_{\mu\nu}$ for whatever analytic continuation we have chosen, evaluate the vector $n$ and boundary metric $\gamma$, plug the result into  (\ref{m}) ( or  (\ref{n}) ) for computing the above result (\ref{q}) and then, finally, either (\ref{pp1}) or (\ref{pp2}). The outcome shall be compared with the results obtained from the minimal surface.

In both analytic continuations we will consider the same background metric we already used for pure AdS, (\ref{f}), dictated by regularity when $t\in[0,2\beta\pi)$ and, being $t-$independent, obviously satisfying any $t-$periodicity conditions we want to ask for any value of $\beta$. For convenience we rescale certain perturbations by factors of $r/\beta$ and by the common factor $R^2/z^2$ and write, using the Fefferman-Graham gauge, 
\begin{equation}\label{i}
ds^2=\frac{R^2}{z^2}( dz^2+(1+\delta\; h_{rr} )dr^2+\frac{r^2}{\beta^2}(1+\delta\; h_{tt} )dt^2+(1+\delta\; h_{22} )dx_2^2+(1+\delta\; h_{33} )dx_3^2
\end{equation}
\[
+2\delta \; \frac{r}{\beta}h_{rt} dr dt+2 \delta\; h_{r2} dr dx_2 +2 \delta \;h_{r3} dr dx_3+2 \delta \;\frac{r}{\beta}h_{t2} dt dx_2+2 \delta \;\frac{r}{\beta} h_{t3} dt dx_3 +2 \delta \;h_{23} dx_2 dx_3) ,
\]
where obviously $h_{\mu\nu}=h_{\mu\nu}(\beta,z,r,t,x_2,x_3)$ and $\delta << 1$. This is the form we will use throwout the paper ( but not in the next section ). The difference will come in the explicit shape for the $h_{\mu\nu}$, and the periodicity conditions they will satisfy. 

\section{First case, keeping $2 \pi \beta$ periodicity }\label{bbb}

\subsection{Minimal area}

As the minimal area computation is done at $\beta=1$, the result should be independent by whichever analytic continuations we are considering when solving for $h_{\mu\nu}$. However in practice, as the equations of motion will be solved using different choices for the Fourier transform, it is more convenient to treat the two cases separately. To compute the perturbation to the minimal area it is better to go back to the Poincare coordinates (\ref{g}). In the Fefferman-Graham gauge ($x^{\mu},x^{\nu}\in\vec{x}=(x^{1},x^{2},x^{3},t_E)$)
\begin{equation}\label{a}
ds^2=\frac{R^2}{z^2}\left( dz^2+g_{\mu\nu}(z,\vec{x})dx^{\mu}dx^{\nu}\right) ,
\end{equation}
\begin{equation}\label{b}
g_{\mu\nu}(z,\vec{x})=\delta_{\mu\nu}+\delta\;h_{\mu\nu}(z,\vec{x}) .
\end{equation}
The equation for $h_{\mu\nu}(z,\vec{x}) $ have been solved ( there in the Lorentzian signature ) at first order in $\delta$ in \cite{Nozaki:2013vta}:
\begin{equation}\label{c}
h_{\mu\nu}(z,\vec{x})=\int d\vec{k} e^{i \vec{k}\cdot\vec{x}}h_{\mu\nu}(z,\vec{k})
\end{equation}
\begin{equation}\label{d}
h_{\mu\nu}(z,\vec{k})=-\frac{8 z^2}{(k_1^2+k_2^2+k_3^2+k_{t_E}^2)} J_2(z i\sqrt{k_1^2+k_2^2+k_3^2+k_{t_E}^2})c_{\mu\nu}(\vec{k})
\end{equation}
( the normalization is to ensure $h_{\mu\nu}(z\rightarrow 0,\vec{k})\rightarrow z^4 c_{\mu\nu}(\vec{k})$ ). As the surface is extremal we need only to compute the variation of the area due to the metric variation, and not the change of shape, that is an higher order effect. Then
\[
\Delta S_A=\frac{1}{4 G_N}\Delta\int dx_2dx_3dz \sqrt{\det(\partial_a x^{\mu}\partial_b x^{\nu}g_{\mu\nu})} \;\;\;\;\;\;  a,b=x_2,x_3,z
\]
and, using (\ref{a}) and (\ref{b}) we obtain
\[
\Delta S_A=\frac{ R^3 \delta}{8 G_N}\int dx_2 dx_3\int_{\epsilon_z}^{L_z}dz \;(\frac{h_{22}(z,0,0,x_2,x_3)+h_{33}(z,0,0,x_2,x_3)}{z^3})
\]
As we will always encounter an integral over $x_2,x_3$ that, being unbounded, should eventually be regularized, both for minimal surfaces and gravitational action, from now on we will simply compare the corresponding integrand $\mathcal{S}_A$ to avoid further issues:
\[
\Delta S_A=\int dx_2 dx_3 \;\Delta \mathcal{S}_A .
\]
Then using (\ref{c}) and (\ref{d}),
\begin{equation}\label{e}
\Delta \mathcal{S}_A=-\int d\vec{k} \; e^{i( k_2 x_2+k_3 x_3)}\frac{ R^3 \delta(c_{22}(\vec{k})+c_{33}(\vec{k}))}{2 G_N (k_1^2+k_2^2+k_3^2+k_{t_E}^2)} ({}_0F_1^R(2,\frac{\epsilon_z^2}{4}(k_1^2+k_2^2+k_3^2+k_{t_E}^2))-{}_0F_1^R(2,\frac{L_z^2}{4}(k_1^2+k_2^2+k_3^2+k_{t_E}^2))) ,
\end{equation}
where ${}_0F_1^R(;a;z)={}_0F_1(;a;z)/\Gamma(a)$ is the regularized version of the confluent Hypergeometric function. Note that we had to introduce a cutoff in $z$ also for large values, $L_z>>1$. 

We can at this point, if we wish, compute the space integral over $x_2,x_3$ of $\Delta\mathcal{S}_A$. As the only $x_2,x_3$ dependence is in the exponential factor, the result is simply proportional to a delta function with $k_2$ and $k_3$ support:
\begin{equation}\label{cca}
\Delta S_A = -\int dk_1 dk_{t_E}\frac{ 2\pi^2 R^3 \delta(c_{22}(k_1,0,0,k_{t_E})+c_{33}(k_1,0,0,k_{t_E}))}{ G_N (k_1^2+k_{t_E}^2)} ({}_0F_1^R(2,\frac{\epsilon_z^2}{4}(k_1^2+k_{t_E}^2))-{}_0F_1^R(2,\frac{L_z^2}{4}(k_1^2+k_{t_E}^2))) .
\end{equation}

\subsection{Gravitational action}

We want to solve the 15 linearized equations of motion for $h_{\mu\nu}$ defined as in (\ref{i}), that are listed in Appendix \ref{B}, while imposing periodicity for $t\rightarrow t+2\pi\beta$ for $\beta\in\mathbb{R}$. The solution looks as follows
\[
h_{\mu\nu}(\beta,z,r,t,x_2,x_3)= \int d\vec{k} \;e^{i(k_1 r \cos(\frac{t}{\beta})+k_{t_E} r \sin(\frac{t}{\beta})+k_2 x_2 +k_3 x_3)}\tilde{h}_{\mu\nu}(\beta,z,\vec{k},t)
\]
\begin{eqnarray}\label{h}
\tilde{h}_{rr}(\beta,z,\vec{k},t)&=&\cos(t/\beta)^2 h_{11}(\beta,z,\vec{k}) +\sin(t/\beta)^2 h_{t_Et_E}(\beta,z,\vec{k})+\sin(2 t/\beta) h_{1t_E}(\beta,z,\vec{k}) \nonumber \\
\tilde{h}_{rt}(\beta,z,\vec{k},t)&=& (\cos(t/\beta)^2- \sin(t/\beta)^2) h_{1t_E}(\beta,z,\vec{k})+\sin(2 t/\beta) (h_{t_E t_E}(\beta,z,\vec{k})-h_{11}(\beta,z,\vec{k}))/2 \nonumber \\
\tilde{h}_{r2}(\beta,z,\vec{k},t)&=& \cos(t/\beta)h_{12}(\beta,z,\vec{k})+\sin(t/\beta)h_{t_E2}(\beta,z,\vec{k}) \nonumber \\
\tilde{h}_{r3}(\beta,z,\vec{k},t)&=&\cos(t/\beta)h_{13}(\beta,z,\vec{k})+\sin(t/\beta)h_{t_E3}(\beta,z,\vec{k}) \nonumber \\
\tilde{h}_{tt}(\beta,z,\vec{k},t)&=&\cos(t/\beta)^2 h_{t_E t_E}(\beta,z,\vec{k})+\sin(t/\beta)^2 h_{11}(\beta,z,\vec{k})-\sin(2 t/\beta) h_{1t_E}(\beta,z,\vec{k})  \nonumber \\
\tilde{h}_{t2}(\beta,z,\vec{k},t)&=&\cos(t/\beta)h_{t_E2}(\beta,z,\vec{k}) -\sin(t/\beta)h_{1 2}(\beta,z,\vec{k})\nonumber \\
\tilde{h}_{t3}(\beta,z,\vec{k},t)&=& \cos(t/\beta)h_{t_E3}(\beta,z,\vec{k}) -\sin(t/\beta)h_{13}(\beta,z,\vec{k}) \nonumber \\
\tilde{h}_{22}(\beta,z,\vec{k})&=& h_{22}(\beta,z,\vec{k}), \;\;\;\tilde{h}_{23}(\beta,z,\vec{k})= h_{23}(\beta,z,\vec{k}), \;\;\;\tilde{h}_{33}(\beta,z,\vec{k})= h_{33}(\beta,z,\vec{k}) , \nonumber \\
\end{eqnarray}
where the $h_{\mu\nu}(\beta,z,\vec{k})$ on the right hand side are the same Fourier coefficients of (\ref{c}) whose explicit expression is (\ref{d}), and we have allowed an additional $\beta$ dependence inside $c_{\mu\nu}(\beta,\vec{k})$. In addition there are five algebraic constraints  \footnote{the same appearing in \cite{Nozaki:2013vta} after the substitutions $k_{t_E}\rightarrow -i w$, $h_{t_Et_E}\rightarrow -h_{tt}$, $h_{1t_E}\rightarrow i h_{1t}$, $h_{t_E2}\rightarrow i h_{t2}$ and $h_{t_E3}\rightarrow i h_{t3}$}:
\begin{eqnarray}\label{h2}
h_{11}&=& -\frac{(k_2^2+k_{t_E}^2)h_{22}+2 k_2 k_3 h_{23}+(k_3^2 +k_{t_E}^2)h_{33}+2 k_1 k_2 h_{12}+2 k_1 k_3 h_{13}}{k_{t_E}^2+k_1^2}\nonumber \\ 
h_{1t_E}&=& \frac{(k_2^2+k_{t_E}^2)k_1 h_{22}+2 k_2 k_3 k_1 h_{23}+(k_3^2 +k_{t_E}^2)k_1h_{33}+ k_2(k_1^2-k_{t_E}^2) h_{12} }{k_{t_E}(k_{t_E}^2+k_1^2)}+\nonumber \\
&+&\frac{k_3(k_1^2-k_{t_E}^2) h_{13}}{k_{t_E}(k_{t_E}^2+k_1^2)}\nonumber \\
h_{t_Et_E}&=& \frac{(k_2^2-k_1^2) h_{22}+2 k_2 k_3 h_{23}+(k_3^2 -k_1^2)h_{33}+2 k_1 k_2 h_{12}+2 k_1 k_3 h_{13}}{k_{t_E}^2+k_1^2}\nonumber \\
h_{t_E2}&=&-\frac{k_2h_{22}+k_3h_{23}+k_1h_{12}}{k_{t_E}} \nonumber \\
h_{t_E3}&=&-\frac{k_3h_{33}+k_2h_{23}+k_1h_{13}}{k_{t_E}}.\nonumber \\ 
\end{eqnarray}
Using the above solution, after a change of coordinates to $u(z,r),\rho(z,r)$, we want to evaluate (\ref{pp1}).  We begin by computing
\[
\int_0^{2\pi} dt \;\partial_{\beta}\left( \Delta  L_{Eucl}^{grav}(\beta,t)\right)_{\beta=1} ,
\]
that is the $\beta$ derivative of (\ref{q}). The explicit expression for the various terms appearing in (\ref{q}) are given in Appendix (\ref{A}) as a function of the perturbation $h_{\mu\nu}$; plugging in the above solution we obtain the required result. We start with the first line of  (\ref{q}) and notice that the two terms just differ by the sign and the replacement $\epsilon_z\leftrightarrow L_z$. Thus we limit ourselves to the computation of the first using the formula for $s_{\rho}$ (\ref{sun}). As we have already seen, the extremes of integration happen to be $\beta$ independent, so that the derivative acts only on the integrand. We have
\begin{equation}\label{uno}
\int_0^{2\pi} dt \int_{\sqrt{\epsilon_r^2+\epsilon_z^2}}^{\sqrt{\epsilon_r^2+\epsilon_z^2}\frac{L_r}{\epsilon_r}}d\rho\;\partial_{\beta}\;s_{\rho}(\beta,u\equiv\frac{\beta\epsilon_z}{\epsilon_r},\rho,t)-(\epsilon_z\leftrightarrow L_z) .
\end{equation}
We are interested in the limit $\epsilon_r\rightarrow 0$. As we do not know how to compute the integral, what we can do is to write the integrand as a series in positive and negative powers of $\rho$ and $u$, compute the integral for each single term and analyse finite and divergent contributions. Then, temporarily ignoring the time integral,
\[
\int_{\sqrt{\epsilon_r^2+\epsilon_z^2}}^{\sqrt{\epsilon_r^2+\epsilon_z^2}\frac{L_r}{\epsilon_r}}d\rho\;\partial_{\beta}\;s_{\rho}(\beta,u \equiv\frac{\beta\epsilon_z}{\epsilon_r},\rho,t)= \hspace{-0,5cm}\sum_{n(\neq -1),k\in \mathbb{Z}}\left(\partial_{\beta}C_{n,k}(\beta,t)+\frac{k}{\beta}C_{n,k}(\beta,t)\right)\left(\frac{\beta\epsilon_z}{\epsilon_r}\right)^k\left[\frac{\rho^{n+1}}{n+1}\right]_{\sqrt{\epsilon_r^2+\epsilon_z^2}}^{\sqrt{\epsilon_r^2+\epsilon_z^2}\frac{L_r}{\epsilon_r}}
\]
\[
+ \sum_{k\in \mathbb{Z}}\left(\partial_{\beta}C_{-1,k}(\beta,t)+\frac{k}{\beta}C_{-1,k}(\beta,t)\right)\left(\frac{\beta\epsilon_z}{\epsilon_r}\right)^k\left[\log(\rho) \right]_{\sqrt{\epsilon_r^2+\epsilon_z^2}}^{\sqrt{\epsilon_r^2+\epsilon_z^2}\frac{L_r}{\epsilon_r}} .
\]
From direct computation we can check that $C_{n,k}=0$ for every $k$ and $n\le 0$ ( no log term ), and for every $n$ and $k\ge 1$. Then terms not vanishing in the $\epsilon_r\rightarrow 0$ limit, either divergent or $\epsilon_r$ independent, can come only from $n>0$ and $k=-(n+1),-n,\dots,-1,0 $. In particular the case with $k=0$ has contribution when $\epsilon_r\rightarrow 0$ from both integration extrema, while for the others values of $k$ only at $\rho=\sqrt{\epsilon_r^2+\epsilon_z^2}\frac{L_r}{\epsilon_r}$. We can compute the $\rho$ and time integral of the $k=0$, $n>0$ term ( zero order in $\epsilon_r$ of the integrand and generic function in $\rho$, as the negative powers are vanishing ), and obtain
\[
-\int d\vec{k} \;e^{i(k_2 x_2+k_3 x_3)}\frac{\delta R^3}{4 G_N \left(k_1^2+k_2^2+k_3^2+k_{t_E}^2\right)}\left(2 c_{t_E t_E}(\beta,\vec{k})-4c_{1 1}(\beta,\vec{k})+\partial_{\beta}c_{t_E t_E}(\beta,\vec{k})+\partial_{\beta}c_{11}(\beta,\vec{k})\right)\cdot
\]
\begin{equation}
\cdot\left({}_0F_1^R(2,\frac{(\epsilon_r^2+\epsilon_z^2)}{4}(k_1^2+k_2^2+k_3^2+k_{t_E}^2))-{}_0F_1^R(2,\frac{L_r^2}{4}\left(1+\frac{\epsilon_z^2}{\epsilon_r^2}\right)(k_1^2+k_2^2+k_3^2+k_{t_E}^2)) \right).
\end{equation}
When $\epsilon_r$ goes to zero the Hypergeometric in the first term simply converges to a function of the only $\epsilon_z$. The other instead is badly divergent in $\epsilon_r$ ( and $L_r$ ), going like $\exp(\sim \epsilon_z\sqrt{k_1^2+k_2^2+k_3^2+k_{t_E}^2} \frac{L_r}{\epsilon_r})$. We disregard this divergence and keep only the finite contribution. 

As we will encounter $\epsilon_r$ singularities throughout the paper let me briefly comment on it. One of the properties of the gravitational solution considered in \cite{Lewkowycz:2013nqa} was to contain only mild singularities whenever $r\rightarrow 0$, that would automatically drop off once computing the entropy $S$; more specifically it was supposed to be possible to constrain the solution such that this behaviour would happen. As here we are presently considering a different periodicity then the one used by Lewkowycz and Maldacena, one would think that the issue arises precisely because of this and it would disappear once the L-M periodic conditions on the gravitational solution are applied. Nonetheless the same divergences will be present in the next section as well, without any clear possibility to constrain the solution so that they would drop when computing the entropy\footnote{ in fact, beside exponentials, we will encounter also polynomial singularities, that could in principle be removed by appropriate counterterms}. The result is that we will limit ourselves to consider the finite part of the entropy, but a better understanding of this issue is important. 

The result of this "ad hoc" renormalization, together with the subtraction of the corresponding $\epsilon_z\rightarrow L_z$ term, is a first contribution to the entropy: 
\[
\Delta\mathcal{S}_1=-\int d\vec{k} \;e^{i(k_2 x_2+k_3 x_3)}\frac{\delta R^3}{4 G_N \left(k_1^2+k_2^2+k_3^2+k_{t_E}^2\right)}\left(2 c_{t_E t_E}(\beta,\vec{k})-4c_{1 1}(\beta,\vec{k})+\partial_{\beta}c_{t_E t_E}(\beta,r\vec{k})+\partial_{\beta}c_{11}(\beta,\vec{k})\right)\cdot
\]
\begin{equation}\label{c1}
\cdot\left({}_0F_1^R(2,\frac{\epsilon_z^2}{4}(k_1^2+k_2^2+k_3^2+k_{t_E}^2))-{}_0F_1^R(2,\frac{L_z^2}{4}(k_1^2+k_2^2+k_3^2+k_{t_E}^2)) \right),
\end{equation}
where we wrote
\[
\Delta S=\int dx_2 dx_3 \;\Delta\mathcal{S}, \;\;\;\;\;\Delta\mathcal{S}=\Delta\mathcal{S}_1+\Delta\mathcal{S}_2+\dots
\]

Then we pass to compute the $n>0$ and $k=-(n+1),-n,\dots,-1 $ coefficients, for each power of $n$ and integrated over time. What happens is that for each one we obtain a divergent term ( negative power in $\epsilon_r$ ), generically $\epsilon_z$ dependent, that we throw away, and a finite contribution always $\epsilon_z$ independent. Thus when we consider the corresponding term with opposite sign after the $\epsilon_z\leftrightarrow L_z$ substitution, all of them simply vanish. 

Let us move to the second line of (\ref{q}). Now the integration extrema are $\beta$ dependent; we can however change integration variable from $u\rightarrow \tilde{u}= u \frac{\epsilon_r}{\beta}$ in such a way to remove the $\beta$ dependence from the integration extrema and make them finite in the limit $\epsilon_r\rightarrow 0$. So we have
\begin{equation}\label{q2}
-\int_0^{2\pi} dt  \int_{\epsilon_z}^{L_z}d\tilde{u}\;\partial_{\beta}\left(\frac{\beta}{\epsilon_r}\;( s_u(\beta,\tilde{u},\epsilon_{\rho}(\beta,\tilde{u}),t)-s_u(\beta,\tilde{u},L_{\rho}(\beta,\tilde{u}),t)) \right) ,
\end{equation}
that more explicitly can be written as
\[
\int_0^{2\pi} dt  \int_{\epsilon_z}^{L_z}\frac{d\tilde{u}}{\epsilon_r}\;\Big( -s_u(\beta,\tilde{u},\sqrt{\epsilon_r^2+\tilde{u}^2},t)+s_u(\beta,\tilde{u},\frac{L_r}{\epsilon_r}\sqrt{\epsilon_r^2+\tilde{u}^2},t) -\tilde{u}\partial_{\tilde{u}}s_u(\beta,\tilde{u},\sqrt{\epsilon_r^2+\tilde{u}^2},t)+
\]
\[
+\tilde{u}\partial_{\tilde{u}}s_u(\beta,\tilde{u},\frac{L_r}{\epsilon_r}\sqrt{\epsilon_r^2+\tilde{u}^2},t)-\beta\partial_{\beta}s_u(\beta,\tilde{u},\sqrt{\epsilon_r^2+\tilde{u}^2},t)
+\beta\partial_{\beta}s_u(\beta,\tilde{u},\frac{L_r}{\epsilon_r}\sqrt{\epsilon_r^2+\tilde{u}^2},t)\Big) .
\]
We start from the case with $\Theta$ term; the above integrand is given by (\ref{srhosn}), transforming to $\tilde{u}$ and plugging in our solution for $h_{\mu\nu}$. In particular direct computation shows that the terms $L_r$ independent vanish in the limit $\epsilon_r\rightarrow 0$ while the ones containing $L_r$ give rise only to the same badly exponentially divergent contribution already encountered. As we have already done we disregard them. 

An analogous story can be told when we have a $\Theta$ term by using (\ref{srhocn}).

We remain with
\[
-\int_{0}^{2\pi}dt\;\Delta L_{Eucl}^{grav}(t)+2\pi \Delta L_{Eucl}^{grav}(2\pi) .
\]
The computation proceeds more or less in the same way, with the only finite term coming from the $\epsilon_r$ independent term in the integrand of (\ref{sun}) ( $k=0$, $n>0$ in the formalism used so far ), and discarding divergences. The result is
\[
\Delta\mathcal{S}_2=\int d\vec{k} \;e^{i(k_2 x_2+k_3 x_3)}\frac{\left(c_{t_E t_E}(\beta,\vec{k})-c_{1 1}(\beta,\vec{k})\right)3\delta R^3}{4 G_N \left(k_1^2+k_2^2+k_3^2+k_{t_E}^2\right)}\cdot
\]
\begin{equation}\label{cc1}
\cdot \left({}_0F_1^R(2,\frac{\epsilon_z^2}{4}(k_1^2+k_2^2+k_3^2+k_{t_E}^2))-{}_0F_1^R(2,\frac{L_z^2}{4}(k_1^2+k_2^2+k_3^2+k_{t_E}^2))\right) .
\end{equation}
As no more contributions are present, we have then achieved an expression for (\ref{pp1}), the finite result given by the sum $\Delta\mathcal{S}=\Delta\mathcal{S}_1+\Delta\mathcal{S}_2$ of (\ref{c1}) and (\ref{cc1}). This is not however the end, as explained in the next section.

\subsection{Restoring $2\pi$ periodicity for $\beta\in\mathbb{Z}$ }

Our metric perturbation $h_{\mu\nu}$ has been constructed in (\ref{h}) to be periodic after a shift of the time variable $t\rightarrow t+2\pi\beta$ for any value of $\beta$. However when $\beta\in\mathbb{Z}$ the present solution does not become automatically periodic for $t\rightarrow t+2\pi$, as we would like. In other words what we would need is to further constrain the $h_{\mu\nu}(z,\vec{k})$ in order to enforce $t\rightarrow t+2\pi$ periodicity as long as $\beta\in\mathbb{Z}$. The solution will then retain by construction only the $t\rightarrow t+2\beta\pi$ periodicity once $\beta\in\mathbb{R}$. 

Let us first show what $h_{\mu\nu}(z,\vec{k})$ should look like in order to have periodicity in $t\rightarrow t+2\pi\beta$ and $t\rightarrow t+2\pi$ when $\beta\in\mathbb{Z}$. To this end let us transform in polar coordinates the Fourier momenta $k_1$ and $k_{t_E}$:
\begin{eqnarray}\label{fmp}
k_1&=& w_r \cos{\frac{\psi}{\beta}} \nonumber \\
k_{t_E}&=& w_r \sin{\frac{\psi}{\beta}} \nonumber \\
\end{eqnarray}
so that
\[
\int dk_1 dk_{t_E}\;e^{i(k_1 x_1 +k_{t_E} t_E)}=\int dw_r\int_{0}^{2\pi\beta} \hspace{-0,3cm}d\psi\;w_r \;e^{i w_r  r\left(\cos{\frac{\psi}{\beta}}\cos{\frac{t}{\beta}}+\sin{\frac{\psi}{\beta}}\sin{\frac{t}{\beta}}\right)}=\int dw_r\int_{0}^{2\pi\beta} \hspace{-0,3cm}d\psi\;w_r \;e^{i w_r  r \cos{\left(\frac{\psi}{\beta}-\frac{t}{\beta}\right)}} .
\]
Fourier transforming $h(\psi,w_r,\beta)$ in this way, will make the resulting function in $t,r$ clearly periodic for $t\rightarrow t+2\pi\beta$. Moreover, if $h(\psi,w_r,\beta)$ is periodic in  $\psi\rightarrow \psi+2\pi$, we can rescale the integration variable $\psi\rightarrow\tilde{\psi}=\psi-2\pi$ in order to absorb also the shift $t\rightarrow t+2\pi$, as long as $\int_{0}^{2\pi\beta}d\psi=\int_{-2\pi}^{2\pi\beta-2\pi}d\tilde{\psi}$. That is we need periodicity for $h(\psi,w_r,\beta)$ also when $\psi\rightarrow \psi+2\pi\beta$, which is compatible with $\psi\rightarrow \psi+2\pi$ only if $\beta\in\mathbb{Z}$. 

Let us implement the solution of $h(\psi,w_r,\beta)=h(\psi-2\pi,w_r,\beta)$ and write it as a function of $k_1,k_{t_E}$. The transformation $\psi\rightarrow \psi-2\pi$ translates in $k_1,k_{t_E}$ variables to
\begin{eqnarray}\label{trsf}
k_{t_E} &\rightarrow & k_{t_E}\cos{\frac{2\pi}{\beta}}-k_1\sin{\frac{2\pi}{\beta}} \nonumber \\
k_{1} &\rightarrow & k_{t_E}\sin{\frac{2\pi}{\beta}}+k_1\cos{\frac{2\pi}{\beta}} .\nonumber \\ 
\end{eqnarray}
A function of $k_1,k_{t_E}$ invariant under (\ref{trsf}) should depend only on $a[k_1,k_{t_E}]\equiv k_1^2+k_{t_E}^2$ and $b[k_1,k_{t_E},\beta]\equiv e^{i n\beta \arctan[k_1,k_{t_E}]}$, $n\in \mathbb{Z}$.\footnote{for $\arctan[k_1,k_{t_E}]$ we mean a function that gives the angle whose tangent is the ratio $k_{t_E}/k_1$, and taking into account the quadrant (  $[0,\pi/2)$ when $k_{t_E}\geq 0,k_1>0$, $[\pi/2,\pi)$ when $k_{t_E}> 0,k_1\leq 0$,  $[\pi,3/2\pi)$ when $k_{t_E}\leq 0,k_1<0$ and $[3/2\pi,2\pi)$ when $k_{t_E}< 0,k_1\geq 0$ ).} Further we rewrite the expressions in (\ref{h}) for $h_{\mu\nu}(\beta,z,r,t,x_2,x_3)$ without the spurious terms in $\sin{(t/\beta)},\cos{(t/\beta)}$ by integrating by parts in the Fourier space,
\begin{eqnarray}\label{h3}
h_{rr}(\beta,z,r,t,x_2,x_3)&=& -\int d\vec{k}\;e^{i(k_1 r \cos(\frac{t}{\beta})+k_{t_E} r \sin(\frac{t}{\beta})+k_2 x_2 +k_3 x_3)}\frac{1}{r^2} (\partial_{k_1}^2 h_{11}(\beta,z,\vec{k}) +\nonumber \\
&+&\partial_{k_{t_E}}^2 h_{t_Et_E}(\beta,z,\vec{k})+2\partial_{k_1}\partial_{k_{t_E}} h_{1t_E}(\beta,z,\vec{k})) \nonumber \\
h_{rt}(\beta,z,r,t,x_2,x_3)&=& -\int d\vec{k}\;e^{i(k_1 r \cos(\frac{t}{\beta})+k_{t_E} r \sin(\frac{t}{\beta})+k_2 x_2 +k_3 x_3)}\frac{1}{r^2}(\partial_{k_1}^2h_{1t_E}(\beta,z,\vec{k})-\nonumber \nonumber \\
&-&\partial_{k_{t_E}}^2h_{1t_E}(\beta,z,\vec{k})+\partial_{k_1}\partial_{k_{t_E}} h_{t_E t_E}(\beta,z,\vec{k})-\partial_{k_1}\partial_{k_{t_E}} h_{11}(\beta,z,\vec{k})) \nonumber \\
etc\dots .\nonumber \\ 
\end{eqnarray}
As we want invariance under (\ref{trsf}) we first require $h_{22}(\beta,z,\vec{k}),h_{23}(\beta,z,\vec{k})$ and $h_{33}(\beta,z,\vec{k})$ to be functions of only $a(k_1,k_{t_E})$ and $b(k_1,k_{t_E},\beta)$. Then we constrain $h_{12}(\beta,z,\vec{k})$ and $h_{13}(\beta,z,\vec{k})$, the only two remaining independent functions according to (\ref{h2}), so that all the expressions on the right hand side of the exponential in (\ref{h3}) are themselves functions of only $a(k_1,k_{t_E})$ and $b(k_1,k_{t_E},\beta)$. That is ( focusing only on the $k_1,k_{t_E}$ variables )
\begin{eqnarray}\label{h4}
\partial_{k_1}^2 h_{11}(k_1,k_{t_E}) +\partial_{k_{t_E}}^2 h_{t_Et_E}(k_1,k_{t_E})+2\partial_{k_1}\partial_{k_{t_E}} h_{1t_E}(k_1,k_{t_E})\hspace{-0.2 cm}&=&\hspace{-0.2 cm} f1\left(a(k_1,k_{t_E}),b(k_1,k_{t_E},\beta)\right)\nonumber \\
\partial_{k_1}^2 h_{1t_E}(k_1,k_{t_E})-\partial_{k_{t_E}}^2h_{1t_E}(k_1,k_{t_E})+\partial_{k_1}\partial_{k_{t_E}} h_{t_E t_E}(k_1,k_{t_E})-\partial_{k_1}\partial_{k_{t_E}} h_{11}(k_1,k_{t_E})\hspace{-0.2 cm}&=& \hspace{-0.2 cm} f2\left(a(k_1,k_{t_E}),b(k_1,k_{t_E},\beta)\right) \nonumber \\
etc\dots .\nonumber \\ 
\end{eqnarray}
For generic $f1$, $f2$ etc... After a lengthy computation we arrive at the, fortunately simple result
\begin{eqnarray}\label{h5}
c_{12}(\beta,\vec{k})&=&-\frac{k_1}{a}\left(k_2c_{22}(a,b,k_2,k_3)+k_3c_{23}(a,b,k_2,k_3)\right)\nonumber \\
c_{13}(\beta,\vec{k})&=&-\frac{k_1}{a}\left(k_3c_{33}(a,b,k_2,k_3)+k_2c_{23}(a,b,k_2,k_3)\right).\nonumber \\ 
\end{eqnarray}
Note that these two functions are not by themselves invariant under (\ref{trsf}) but they transform in such a way to make each of the $h_{\mu\nu}$ given by (\ref{h}),(\ref{h2}), (\ref{h3}) and (\ref{h5}) ultimately periodic in $t\rightarrow t+2\pi$ ( for integer $\beta$ ). Thus we ended up with only three independent functions: $h_{22}$, $h_{23}$ and $h_{33}$. In particular $c_{tt}(\beta,\vec{k})$ and $c_{11}(\beta,\vec{k})$ now are
\begin{eqnarray}\label{h6}
c_{tt}(\beta,\vec{k})&=&\frac{1}{a^2}\Big((k_2^2(k_{t_E}^2-k_1^2)-k_1^2 a)c_{22}(a,b,k_2,k_3)+2k_2 k_3(k_{t_E}^2-k_1^2)c_{23}(a,b,k_2,k_3)+\nonumber \\
&+&(k_3^2(k_{t_E}^2-k_1^2)-k_1^2 a)c_{33}(a,b,k_2,k_3)\Big)\nonumber \\
c_{11}(\beta,\vec{k})&=&-\frac{1}{a^2}\Big((k_2^2(k_{t_E}^2-k_1^2)+k_{t_E}^2 a)c_{22}(a,b,k_2,k_3)+2k_2 k_3(k_{t_E}^2-k_1^2)c_{23}(a,b,k_2,k_3)+\nonumber \\
&+&(k_3^2(k_{t_E}^2-k_1^2)+k_{t_E}^2 a)c_{33}(a,b,k_2,k_3)\Big).\nonumber \\ 
\end{eqnarray}
We have chosen to avoid any direct dependence on $\beta$ inside $c_{22},\;c_{23}$ and $c_{33}$, but only through $b(k_1,k_{t_E},\beta)$. This will turn out to be the correct prescription. We can insert the result in (\ref{c1}) and, after a change of integration variables from $k_1,k_{t_E}$ to $a,b$, an integration by part in $b$, and a change back to $k_1,k_{t_E}$, we obtain the result for $\Delta\mathcal{S}_1$ to be transformed from (\ref{c1}) to:
\[
\Delta\mathcal{S}_1=-\int d\vec{k} \frac{e^{i(k_2 x_2+k_3 x_3)}\delta R^3}{4 G_N \left(k_1^2+k_2^2+k_3^2+k_{t_E}^2\right)\left(k_1^2+k_{t_E}^2\right)^2}\Big( (5 k_{t_E}^4+4 k_{t_E}^2k_1^2-k_1^4+6 k_2^2(k_{t_E}^2-k_1^2))c_{22}(a,b,k_2,k_3)+
\]
\[
+(5 k_{t_E}^4+4 k_{t_E}^2k_1^2-k_1^4+6 k_3^2(k_{t_E}^2-k_1^2))c_{33}(a,b,k_2,k_3)+12 k_2 k_3 (k_{t_E}^2-k_1^2) c_{23}(a,b,k_2,k_3)\Big)\cdot
\]
\begin{equation}\label{c3}
\cdot \left({}_0F_1^R(2,\frac{\epsilon_z^2}{4}(k_1^2+k_2^2+k_3^2+k_{t_E}^2))-{}_0F_1^R(2,\frac{L_z^2}{4}(k_1^2+k_2^2+k_3^2+k_{t_E}^2))\right) ,
\end{equation}
and analogously for $\Delta\mathcal{S}_2$:
\[
\Delta\mathcal{S}_2=\int d\vec{k} \frac{e^{i(k_2 x_2+k_3 x_3)} 3\delta R^3 (k_{t_E}^2-k_1^2)}{4 G_N \left(k_1^2+k_2^2+k_3^2+k_{t_E}^2\right)\left(k_1^2+k_{t_E}^2\right)^2}\Big( (2 k_2^2+(k_1^2+k_{t_E}^2))c_{22}(a,b,k_2,k_3)+
\]
\[
(2 k_3^2+(k_1^2+k_{t_E}^2))c_{33}(a,b,k_2,k_3)+4 k_2 k_3c_{23}(a,b,k_2,k_3)\Big)\cdot
\]
\begin{equation}\label{cc3}
\cdot\left({}_0F_1^R(2,\frac{\epsilon_z^2}{4}(k_1^2+k_2^2+k_3^2+k_{t_E}^2))-{}_0F_1^R(2,\frac{L_z^2}{4}(k_1^2+k_2^2+k_3^2+k_{t_E}^2))\right) .
\end{equation}
Their sum  gives 
\[
\Delta\mathcal{S}=\Delta\mathcal{S}_1+\Delta\mathcal{S}_2=-\int d\vec{k}\;e^{i(k_2 x_2+k_3 x_3)}\frac{ R^3 \delta(c_{22}(a,b,k_2,k_3)+c_{33}(a,b,k_2,k_3))}{2 G_N (k_1^2+k_2^2+k_3^2+k_{t_E}^2)}\cdot
\]
\begin{equation}\label{cct}
\cdot ({}_0F_1^R(2,\frac{\epsilon_z^2}{4}(k_1^2+k_2^2+k_3^2+k_{t_E}^2))-{}_0F_1^R(2,\frac{L_z^2}{4}(k_1^2+k_2^2+k_3^2+k_{t_E}^2))) ,
\end{equation}
and exactly matches (\ref{e}) \footnote{as long as we restrict the $k_1,k_{t_E}$ dependence to be through the variables $a,b$}. This result is surprising as, having different periodic conditions then in \cite{Lewkowycz:2013nqa}, we would have expected disagreement. This may be an effect of being at first order in $\delta$, somehow mimicking the corresponding boundary situation we encountered when using (\ref{dueb}), or it may be the sign that multiple choices for the analytic continuation can be done if the goal is to compute the entropy. This last possibility does not clash with the construction of Lewkowycz and Maldacena, in fact it even reinforces it as different choices lead to the same result, but it raises the issue of unicity of the analytic continuation and how to physically single out one from another. 

\section{Second case, keeping $2 \pi$ periodicity }\label{ccc}

\subsection{Gravitational action}

Now we want to solve the equations of motion for $h_{\mu\nu}$ so that the solution remains explicitly periodic in $t\rightarrow t+2\pi$ ( but not in $t\rightarrow t+2\pi\beta$ ) after analytic continuation of $\beta$, with the perturbations conveniently chosen as in (\ref{i}). To this end we Fourier transform $h_{\mu\nu}$ as follows
\begin{equation}
h_{\mu\nu}(\beta,z,r,t,x_2,x_3)= \sum_{n_t\in \mathbb{Z}}\int dk_2 dk_3 e^{i(n_t t +k_2 x_2 +k_3 x_3)}h_{\mu\nu}(\beta,z,r,n_t,k_2,k_3) .
\end{equation}
Plugging the above form into the expressions of Appendix \ref{B}, the resulting equations are quite complicated to solve as they are intrinsically coupled between different perturbations. However they simplify if we restrict to the case $h_{rr}=h_{tt}$. As can be seen playing with the equations this implies $\partial_t h_{rt}=0$. To compute the action we do not need to solve all the equations as  (\ref{sun}), (\ref{srhosn}) and (\ref{srhocn}) contain only $h_{rr}$ and $h_{tt}$ ( equal by assumption ). Their equation reduces to (  $h=h_{tt}=h_{rr}$ but it is valid also for $h_{22},h_{33}$ )
\begin{equation}\label{l}
(\beta^2 n_t^2+(k_2^2+k_3^2)r^2)z\; h(z,r)=r z \partial_{r}h(z,r)+r^2(z\partial_{r}^2h(z,r)-3\partial_{z}h(z,r)+z\partial_{z}^2h(z,r)) ,
\end{equation}
where for simplicity we have dropped the dependence of $h$ by $n_t,k_2$ and $k_3$. Applying the ansatz
\[
h(\beta,z,r,n_t,k_2,k_3)=f(\beta,r,n_t)g(\beta,z,n_t,k_2,k_3)
\]
we can decouple the equation in $z$ by the one in $r$ requiring $f(\beta,r,n_t)$ to satisfy
\[
\beta^2 n_t^2 f(\beta,r,n_t)=r\partial_{r}f(\beta,r,n_t)+r^2\partial_{r}^2 f(\beta,r,n_t) .
\]
Then the equation in $z$ becomes
\[
(k_2^2+k_3^2) z\; g(\beta,z,n_t,k_2,k_3)=-3\partial_{z}g(\beta,z,n_t,k_2,k_3)+z\partial_{z}^2 g(\beta,z,n_t,k_2,k_3) ,
\]
the solutions are respectively ( with $g(\beta,z\rightarrow 0,n_t,k_2,k_3)\rightarrow z^4 c(\beta,n_t,k_2,k_3)$ )
\begin{eqnarray}
f(\beta,r,n_t)&=&q_1\cosh(\beta n_t\log(r))+i q_2\sinh(\beta n_t\log(r)) \nonumber \\
g(\beta,z,n_t,k_2,k_3)&=&-\frac{8 z^2}{(k_2^2+k_3^2)} J_2(z i\sqrt{k_2^2+k_3^2})c(\beta,n_t,k_2,k_3). \nonumber \\ 
\end{eqnarray}
At this point both $q_1$ and $q_2$ are free parameters to be correctly normalized, that may depend on $n_t$ and $\beta$. We choose them in order to select solutions with certain asymptotic behaviours; two obvious choices are
\begin{equation}\label{p}
q_2(n_t)=\pm i q_1(n_t)\left(-\Theta(n_t)+\Theta(-n_t) \right) .
\end{equation}
The first ( plus ) making $f(\beta,r,n_t)$ finite in the limit $r\rightarrow 0$, the second ( minus ) in the limit $r\rightarrow\infty$. When $n_t=0$ the asymptotic behaviour in $z\rightarrow 0$ is $r-$independent, so we fix $q_1(0)=1$ in order to have $h(\beta,z\rightarrow 0,r,0,k_2,k_3)\rightarrow  z^4 c(\beta,0,k_2,k_3)$.

Given this solution we want to evaluate (\ref{pp2}). The procedure is analogous as in the previous sections, only with different perturbations involved and, as we are in the case $h_{rr}=h_{tt}$, expressions (\ref{sun}), (\ref{srhosn}) and (\ref{srhocn}) further simplify.

The computation of   (\ref{q2}), both when we do and do not have a $\Theta$ term, gives the same vanishing plus exponentially divergent results, when the $\epsilon_r$ and $L_r$ cutoffs are removed. Thus no finite contribution. Also (\ref{uno}) works similarly and the only finite term that we obtain comes from the order zero in the $\epsilon_r$ expansion of the integrand. As in doing the time integral between $[0,2\pi]$ $e^{i n_t t}$ gives always zero unless $n_t=0$, the choice between the two asymptotic solutions of (\ref{p}) is irrelevant, with $q_2(0)=0$. Then the result is ( using $c_{tt}=c_{rr}=-\frac{1}{2}(c_{22}+c_{33})$ that comes from the first equation of (\ref{aleq}) )
\[
\Delta\mathcal{S}=-\int dk_2dk_3\;\left( c_{22}(1,0,k_2,k_3)+c_{33}(1,0,k_2,k_3)-\partial_{\beta}c_{22}(\beta,0,k_2,k_3)|_{\beta=1}-\partial_{\beta}c_{33}(\beta,0,k_2,k_3)|_{\beta=1}\right)\cdot
\]
\begin{equation}\label{res}
\cdot\frac{ e^{i(k_2 x_2+k_3 x_3)}R^3 \delta}{4 G_N \left(k_2^2+k_3^2\right)}\left({}_0F_1^R(2,\frac{\epsilon_z^2}{4}(k_2^2+k_3^2))-{}_0F_1^R(2,\frac{L_z^2}{4}(k_2^2+k_3^2))\right).
\end{equation}

\subsection{Minimal area}

For comparison with the above result we have to compute the minimal area surface when the perturbation $h_{\mu\nu}$ has been Fourier transformed as in the above section, with the position of the minimal area surface translated from $x_1=t_E=0$ into $t=0,r=\epsilon_r$:
\[
\Delta S_A=\frac{ \delta R^3}{8 G_N}\int dx_2 dx_3\int_{\epsilon_z}^{L_z}dz (\frac{h_{22}(z,\epsilon_r,0,x2,x3)+h_{33}(z,\epsilon_r,0,x2,x3)}{z^3})
\]
and
\begin{equation}\label{o}
\Delta\mathcal{S}_A=-\int dk_2dk_3\;e^{i( k_2 x_2+k_3 x_3)}\frac{ (c_{22}(0,k_2,k_3,1)+c_{33}(0,k_2,k_3,1))R^3  \delta}{2 G_N (k_2^2+k_3^2)} ({}_0F_1^R(2,\frac{\epsilon_z^2}{4}(k_2^2+k_3^2))-{}_0F_1^R(2,\frac{L_z^2}{4}(k_2^2+k_3^2))) ,
\end{equation}
where we have chosen the first solution of equation (\ref{p}) and the contribution coming only from the $n_t=0$ sector, the rest being infinitesimal in the limit $\epsilon_r\rightarrow 0$.

This coincides with (\ref{res}) if we impose the simple condition on $c_{22},c_{33}$: 
\begin{equation}\label{hh6}
\partial_{\beta}c(\beta,0,k_2,k_3)|_{\beta=1}=-c(1,0,k_2,k_3).
\end{equation} 

If we wish we can again perform the integration over $x_2,x_3$ to obtain:
\begin{equation}\label{ops}
\Delta S_A=\frac{ \delta R^3 \pi^2(L_z^2-\epsilon_z^2)(c_{22}(0,0,0,1)+c_{33}(0,0,0,1))  }{4 G_N }.
\end{equation}

\section{Conclusions}

What we have done in this paper is to consider two different possible analytic continuations in $\beta$ of the holographic counterpart of $Tr\left[\left(\mathcal{T}e^{-\int_0^{2\pi}H(t)}\right)^{\beta}\right]$. The guideline for choosing them has been what periodicity to keep for the gravitational solution, out of the two we should impose when $\beta\in\mathbb{Z}$. One possible choice is the one used by Lewkowycz and Maldacena, that allows for a proof of the Ryu-Takayanagi formula, the other is essentially its complement. As the choice done in \cite{Lewkowycz:2013nqa} is an assumption, the idea was to bring additional evidence from a first order computation; instead we have found that both solutions work correctly at first order, reproducing the corresponding result obtained from computing the minimal area. Thus it may in principle be possible to consider additional different, more involved gravitational metrics and to expect some of them to provide the correct result at first order. This is somehow reminiscent of the corresponding boundary problem, where the wrong choice (\ref{dueb}) could nonetheless reproduce the right first order result. An obvious possibility for better understanding the issue would be to go to second order in $\delta$. This however seems impractical, both because of the complication in solving the Einstein's equations and as, at second order, the minimal surface shape should be backreacted. Nonetheless some results appeared recently at second order in \cite{He:2014lfa} so, perhaps, an attempt to go to second order may be worth. In general so far how to provide a more robust understanding for the holographic dual of $Tr[\rho^{\beta}]$ for real $\beta$ is still an open problem.

Besides the above story two additional considerations arise: one being the divergences in $\epsilon_r$ that we had to throw away when performing the computation. The second is the choice for the $\beta$ dependence inside $c_{\mu\nu}(\beta,k)$ that we have made in order to reproduce the correct result from the minimal area. Concerning the divergences we do not have much to say; we solved Einstein's equations at first order for the perturbation of a regular background, but the entropy evaluated from the gravitational action presents divergencies, in contrast with the general analysis in \cite{Lewkowycz:2013nqa}. Further these divergencies appear both as negative powers of $\epsilon_r$, that may be regularized after an appropriate choice of counterterms ( see for example \cite{Karch:2014ufa} ), and as exponentials that are clearly more problematic. A possibility would be to constrain the solution in order to make the entropy finite, but how to do so is not clear. Besides this fact, the finite part of the computation correctly agrees with the minimal area result, for both analytic continuations, only if we assume a certain $\beta$ dependence of the Fourier coefficients, leading respectively to the discussion below equation (\ref{h6}) and to equation (\ref{hh6}). Respecting these additional constrains is not a problem in practice, but their very presence is a novelty.

Finally we treated in detail the issue of cutoffs and coordinates in AdS. This may seem a technical problem but in fact most of the work is directly connected with it: the $(z,r)$ to $(u,\rho)$ transformation permits to correctly compute the entropy and sets the cutoff in $(u,\rho)$ coordinates and their relationship with the ones in $(z,r)$; these then fix the boundaries on which $\Delta S^{grav}_{Eucl}$ lives ( making it non zero ), make its $\beta$ dependence non trivial and, by sending $\epsilon_r\rightarrow 0$, directly implement the restriction of the entropy to the region in AdS where the time circle shrinks. As most of these issues were so far not explicitly covered in the literature, at least in practical examples, we felt compelling a more precise discussion.

The generalized gravitational entropy is not only a useful tool for providing a proof of the Ryu-Takayanagi formula, but it also allows to extend the holographic description of entanglement entropy beyond the original range of applicability; for example for studying bulk quantum corrections \cite{Engelhardt:2014gca} and \cite{Faulkner:2013ana}, or higher derivatives extensions of the gravity theory, for instance \cite{Bhattacharyya:2013jma} \cite{Bhattacharyya:2014yga} \cite{Bhattacharyya:2013gra}  \cite{Camps:2013zua}  \cite{Dong:2013qoa} and \cite{Miao:2014nxa}. This last case case in particular is interesting, as we do not know of any functional such that its minimization would provide a generalization of the Ryu-Takayanagi formula for a bulk theory with generic higher curvature terms. Still using the generalized gravitational entropy it has been possible to construct some formulas for the holographic computation of the entanglement entropy. The technique developed in the present paper may then be extended to some of these theories, for example to compute holographically the reaction and dynamics of the entanglement entropy to sudden changes of parameters, known as quantum quenches.

\section*{Acknowledgements}

I would like to thank Aitor Lewkowycz, Tadashi Takayanagi, Masaki Shigemori and Diego Trancanelli for suggestions and comments on the paper. Further the result of Appendix \ref{C} has been taken from Shigemori's notes, although the proof here is different, and the first motivation for the present work has sourced from several discussions with him. This work was founded by FAPESP fellowship 2013/10460-9.

\appendix
\section{Proof of the first order relationship between $\mathcal{H}$ and $H$}\label{C}

We want to prove the equation
\begin{equation}\label{ham2}
Tr \left[\mathcal{H} e^{-\mathcal{H}}\right]_{order \; O(\delta)}=Tr \left[\mathcal{T}\left(\int_{0}^{2\pi}dt\;H(t) e^{-\int_{0}^{2\pi}d\tilde{t}\;H(\tilde{t})}\right)\right]_{order \;O(\delta)} ,
\end{equation}
where
\[
e^{-\mathcal{H}}\equiv e^{-\int_0^{2\pi}H(t)} \;\;\;\;\; H(t+2\pi)=H(t) ,
\]
and
\[
H(t)=H_0+\delta \;h(t)  \;\;\;\;\; \delta<<1 .
\]
At order zero in $\delta$ the relationship is trivial as, given (\ref{magnus})
\[
Tr \left[\mathcal{H} e^{-\mathcal{H}}\right]_{order \; O(\delta^0)}=Tr \left[2\pi H_0 e^{-2\pi H_0}\right]=Tr \left[\int_{0}^{2\pi} dt\;e^{-(2\pi-t) H_0} H_0 \;e^{-t H_0}\right]
\]
and the right hand side is just the $O(\delta^0)$ term of 
\begin{equation}\label{exp}
Tr \left[\int_{0}^{2\pi}dt\;\mathcal{T}\left(e^{-\int_{t}^{2\pi}d\tilde{t}\;H(\tilde{t})}\right)\;H(t) \mathcal{T}\left(e^{-\int_{0}^{t}d\tilde{t}\;H(\tilde{t})}\right)\right]=Tr \left[\mathcal{T}\left(\int_{0}^{2\pi}dt\;H(t) e^{-\int_{0}^{2\pi}d\tilde{t}\;H(\tilde{t})}\right)\right] .
\end{equation}

At first order in $\delta$ we should work out two additional terms, respectively  when $\delta h(t) $ is picked up either from the $\mathcal{H}$ "downstairs" in (\ref{ham2}) or from $e^{-\mathcal{H}}$. Using  (\ref{magnus}) the first one is
\[
 Tr \left[ \delta\left(\int_0^{2\pi}dt_1\;h(t_1)-\frac{1}{2}\int_0^{2\pi}dt_1\int_0^{t_1}dt_2 \;(h(t_1) H_0-H_0 h(t_1) +H_0 h(t_2)-h(t_2) H_0) +\dots\right)e^{-2\pi H_0}\right]= 
\]
\[
=Tr \left[\delta \int_0^{2\pi}dt\;h(t)e^{-2\pi H_0}\right]=Tr \left[\delta \int_0^{2\pi}dt\;e^{-(2\pi-t) H_0}h(t)\;e^{-t H_0}\right] ,
\]

and the cyclicity of the trace has been used in the first equality to kill all the terms containing a commutator ( with $e^{-2\pi H_0}$ and $H_0$ obviously commuting ). Again the final expression is what we would obtain from (\ref{exp}) at order $O(\delta)$, picking up $h(t)$ from the Hamiltonian below. The final piece is then 
\[
Tr \left[2\pi H_0 \left(e^{-\mathcal{H}}\right)_{O(\delta)}\right]=Tr\left[ 2\pi H_0 \mathcal{T}\left( e^{-\int_0^{2\pi}H(t)}\right)_{O(\delta)}\right]=
\]
\[
=Tr\left[2\pi H_0 \mathcal{T}\left( e^{-\int_t^{2\pi}d\tilde{t}H(\tilde{t})}\right)_{O(\delta^0)} \mathcal{T}\left( e^{-\int_0^{t}d\tilde{t}H(\tilde{t})}\right)_{O(\delta)}\right]+Tr\left[2\pi H_0 \mathcal{T}\left( e^{-\int_t^{2\pi}d\tilde{t}H(\tilde{t})}\right)_{O(\delta)} \mathcal{T}\left( e^{-\int_0^{t}d\tilde{t}H(\tilde{t})}\right)_{O(\delta^0)}\right] ,
\]
and again using the cyclicity of the trace and the fact that $H_0$ and $\mathcal{T}\left( e^{-\int_0^{t}d\tilde{t}H(\tilde{t})}\right)_{O(\delta^0)}$ commute we can rewrite the above expression as the $O(\delta)$ order of 
\[
Tr\left[\int_{0}^{2\pi}dt\;\mathcal{T}\left( e^{-\int_t^{2\pi}d\tilde{t}H(\tilde{t})}\right) H_0\mathcal{T}\left( e^{-\int_0^{t}d\tilde{t}H(\tilde{t})}\right)\right] ,
\]
which is again trivially what we were missing of (\ref{exp})$_{O(\delta)}$.

\section{Equations of motion}\label{B}

Here we list the 15 equations of motion from the independent components of the Einstein tensor ( plus cosmological constant ), expanded at first order in $\delta$. Five equations involve only derivatives with respect to $r,t,x_2$ and $x_3$, and so after Fourier transforming these coordinates they become algebraic. The remaining ten instead are differential equations in $z$. 

As the Fourier transformation involving coordinates $r,t$ will be different depending on the analytic continuation we want to achieve, in this appendix we Fourier transform the $h_{\mu\nu}$ functions only in $x_2,x_3$:
\[
h_{\mu\nu}(x_2,x_3)=\int dk_2dk_3 e^{i k_2 x_2+i k_3 x_3}h_{\mu\nu}(k_2,k_3) .
\]
Further the form of the equations is simpler if expressed in $z,r$ coordinates. After some rearranging the five "algebraic" equations are
\begin{eqnarray}\label{aleq}
h_{22}+h_{33}+h_{rr}+h_{tt}&=&0  \\ \nonumber 
i \;k_2\; r\; h_{r2}+i\; k_3\; r\; h_{r3}+h_{rr}-h_{tt}+\beta\; \partial_t h_{rt} +r \;\partial_r h_{rr} &=&0  \\ \nonumber 
2\; i\; h_{rt} -k_2\; r\; h_{t2}-k_3\;r\; h_{t3} +i\;\beta\;\partial_t h_{tt}+i \;r \;\partial_r h_{rt} &=& 0  \\ \nonumber 
-k_2\; r\; h_{22}-k_3\; r\; h_{23}+i\left( h_{r2}+\beta\;\partial_t h_{t2}+r\;\partial_r h_{r2}\right)&=& 0  \\ \nonumber 
-k_3\; r\; h_{33}-k_2\; r\; h_{23}+i\left( h_{r3}+\beta\;\partial_t h_{t3}+r\;\partial_r h_{r3}\right)&=& 0 . \\  \nonumber 
\end{eqnarray}
And the ten "differential":
\begin{eqnarray}\label{difeq}
r^2 \left(k_2^2+k_3^2 \right)h_{22}&=&\beta^2\partial_t^2 h_{22}+r\partial_r h_{22}+r^2\left( \partial_r^2 h_{22}-3/z\partial_z h_{22}+\partial_z^2 h_{22}\right)\\ \nonumber 
r^2 \left(k_2^2+k_3^2 \right)h_{23}&=&\beta^2\partial_t^2 h_{23}+r\partial_r h_{23}+r^2\left( \partial_r^2 h_{23}-3/z\partial_z h_{23}+\partial_z^2 h_{23}\right)\\ \nonumber 
r^2 \left(k_2^2+k_3^2 \right)h_{33}&=&\beta^2\partial_t^2 h_{33}+r\partial_r h_{33}+r^2\left( \partial_r^2 h_{33}-3/z\partial_z h_{33}+\partial_z^2 h_{33}\right)\\  \nonumber  
r^2 \left(k_2^2+k_3^2 \right)h_{rr}&=&\beta^2\partial_t^2 h_{rr}-2 h_{rr}+2 h_{tt}-4\beta\partial_t h_{rt}+r\partial_r h_{rr}+ r^2\left( \partial_r^2 h_{rr}-3/z\partial_z h_{rr}+\partial_z^2 h_{rr}\right)  \\ \nonumber 
r^2 \left(k_2^2+k_3^2 \right)h_{rt}&=&\beta^2\partial_t^2 h_{rt}-4 h_{rt}+2 \beta\partial_t h_{rr}-2\beta\partial_t h_{tt}+r\partial_r h_{rt}+r^2\left( \partial_r^2 h_{rt}-3/z\partial_z h_{rt}+\partial_z^2 h_{rt}\right)\\ \nonumber 
r^2 \left(k_2^2+k_3^2 \right)h_{r2}&=&\beta^2\partial_t^2 h_{r2}-h_{r2}-2\beta\partial_t h_{t2}+r\partial_r h_{r2}+r^2\left( \partial_r^2 h_{r2}-3/z\partial_z h_{r2}+\partial_z^2 h_{r2}\right) \\ \nonumber 
r^2 \left(k_2^2+k_3^2 \right)h_{r3}&=&\beta^2\partial_t^2 h_{r3}-h_{r3}-2\beta\partial_t h_{t3}+r\partial_r h_{r3}+r^2\left( \partial_r^2 h_{r3}-3/z\partial_z h_{r3}+\partial_z^2 h_{r3}\right) \\ \nonumber 
r^2 \left(k_2^2+k_3^2 \right)h_{tt}&=&\beta^2\partial_t^2 h_{tt}-2 h_{tt}+2 h_{rr}+4\beta\partial_t h_{rt}+r\partial_r h_{tt}+r^2\left( \partial_r^2 h_{tt}-3/z\partial_z h_{tt}+\partial_z^2 h_{tt}\right) \\ \nonumber 
r^2 \left(k_2^2+k_3^2 \right)h_{t2}&=&\beta^2\partial_t^2 h_{t2}-h_{t2}+2\beta\partial_t h_{r2}+r\partial_r h_{t2}+r^2\left( \partial_r^2 h_{t2}-3/z\partial_z h_{t2}+\partial_z^2 h_{t2}\right) \\ \nonumber 
r^2 \left(k_2^2+k_3^2 \right)h_{t3}&=&\beta^2\partial_t^2 h_{t3}-h_{t3}+2\beta\partial_t h_{r3}+r\partial_r h_{t3}+r^2\left( \partial_r^2 h_{t3}-3/z\partial_z h_{t3}+\partial_z^2 h_{t3}\right). \\ \nonumber 
\end{eqnarray}

\section{Some integrand expressions}\label{A}

Let us explicitly evaluate the integrands of (\ref{q}) for perturbations $h_{\mu\nu}$ defined as in (\ref{i}). This appendix does not distinguish between the two possible periodicities after analytic continuation of $\beta$. Further we will work in $u(z,r),\rho(z,r)$ coordinates, as these are the ones to be used in order to naturally implement the cutoff procedure. We first need the unit normal outward pointing vector $n$, for the four boundaries, at first order in $\delta$. The expressions are 
\[
n_{\rho}=\left(-\frac{\rho u}{R\sqrt{\beta^2+u^2}}+\frac{\delta \beta^2 \rho \;u\; h_{rr}(\rho,u,\dots) }{2 R (\beta^2+u^2)^{3/2}}\right)\partial_{\rho} |_{\rho=\epsilon_{\rho},L_{\rho}}   ,
\]
and 
\[
n_{u}=\left(-\frac{u\sqrt{\beta^2+u^2}}{R \beta}+\frac{\delta u^3\; h_{rr}(\rho,u,\dots) }{2 R \beta \sqrt{\beta^2+u^2}}\right)\partial_{u} |_{u=\epsilon_u,L_u}   .
\]
Note that we have decided for simplicity, to keep the same sign for the vector for both the UV and IR cutoff, and include the relative minus sign directly in the expression (\ref{q}) ( which is odd in $n$ ). Given these expressions the results are
\[
s_{\rho}(\beta,u,\rho,t)|_{u=cutoff}=-\frac{\delta R^3}{16 \pi \beta G_N  \rho^3 u^4}\int dx_2 dx_3\;\Big( \left( 3 \beta^4+4\beta^2u^2+u^4\right)(h_{22}+h_{33})+\left( 3 \beta^4+\beta^2u^2+u^4\right)h_{rr}
\]
\begin{equation}\label{su}
+\left( 3 \beta^4+3\beta^2u^2\right)h_{tt}+2\beta\rho u^2\sqrt{\beta^2+u^2}(\partial_{x_2}h_{r2}+\partial_{x_3}h_{r3})+2\beta u^2\left(\beta^2+ u^2\right)\partial_{t}h_{rt}+2\beta^2\rho u^2\partial_{\rho}h_{rr}\Big)
\end{equation}
and 
 \[
s_{u}(\beta,u,\rho,t)|_{\rho=cutoff}=-\frac{\delta \beta R^3}{16 \pi G_N  \rho^2 u^3 \left(\beta^2+u^2\right)}\int dx_2 dx_3\;\Big( \left(\beta^2+u^2\right)(h_{22}+h_{33}+2h_{tt})-8\beta^2h_{rr}
\]
\begin{equation}\label{srhoc}
-2\beta\rho\sqrt{\beta^2+u^2}(\partial_{x_2}h_{r2}+\partial_{x_3}h_{r3})-2\beta(\beta^2+u^2)\partial_{t}h_{rt}-\beta^2\rho\partial_{\rho}h_{rr}\Big)
\end{equation}
or, without the $\Theta$ term,
\[
s_{u}(\beta,u,\rho,t)|_{\rho=cutoff}=\frac{\delta \beta R^3}{16 \pi G_N  \rho^2 u^3 \left(\beta^2+u^2\right)}\int dx_2 dx_3\;\Big(\left(\beta^2+u^2\right)(h_{22}+h_{33})+2\left(2\beta^2+u^2\right)h_{rr}
\]
\begin{equation}\label{srhos}
+2\beta\rho\sqrt{\beta^2+u^2}(\partial_{x_2}h_{r2}+\partial_{x_3}h_{r3})+2\beta(\beta^2+u^2)\partial_{t}h_{rt}-\rho\left(\beta^2+u^2\right)(\partial_{\rho}h_{22}+\partial_{\rho}h_{33}+\partial_{\rho}h_{tt})\Big) .
\end{equation}

We can simplify these expressions using the equations of motion, essentially subtracting terms proportionals to the first two equations of (\ref{aleq}) ( rewritten in $\rho,u$ variables and Fourier transformed back into functions of $x_2,x_3$ ), and reduce them to respectively
\begin{equation}\label{sun}
s_{\rho}(\beta,u,\rho,t)|_{u=cutoff}=-\frac{\delta R^3}{16 \pi \beta G_N  \rho^3 u^2}\int dx_2 dx_3\;\left( -(5\beta^2+2 u^2)h_{rr}+(\beta^2+ u^2)(h_{tt}+2 u \partial_u h_{rr})\right) ,
\end{equation}
\begin{equation}\label{srhosn}
s_{u}(\beta,u,\rho,t)|_{\rho=cutoff}=\frac{\delta \beta R^3}{16 \pi G_N  \rho^2 u^3 \left( \beta^2+u^2\right)}\int dx_2 dx_3\;\big( (7\beta^2- u^2)h_{rr}+(\beta^2+u^2)h_{tt}+
\end{equation}
\[
+2 u(\beta^2+u^2)\partial_u h_{rr}-\beta^2\rho\partial_{\rho}h_{rr}\big)
\]
and
\begin{equation}\label{srhocn}
s_{u}(\beta,u,\rho,t)|_{\rho=cutoff}=-\frac{\delta \beta R^3}{16 \pi G_N  \rho^2 u^3 \left(\beta^2+u^2\right)}\int dx_2 dx_3\;\left( - (3\beta^2+ u^2)h_{rr}- (\beta^2+ u^2)h_{tt}+\beta^2\rho\partial_{\rho}h_{rr}\right) ,
\end{equation}
this last one without the $\Theta$ term.

\end{document}